\documentclass[showpacs,amsmath,amssymb,aps,prd, preprint,groupedaddress,superscriptaddress,nofootinbib]{revtex4-1} % COLOCADO POR MIM
\usepackage[utf8]{inputenc}
\usepackage{graphicx}
\usepackage{tensor}
\usepackage{bm}

\newcommand{\e}[1]{\tensor{e}{#1}}

\newcommand{\R}[1]{\tensor{\mathring{R}}{#1}} %eu
\newcommand{\rcd}[1]{\tensor{\overset{\circ}{\nabla}}{#1}}%eu
\newcommand{\connection}[1]{\tensor{\Gamma}{#1}}%eu
\newcommand{\chr}[1]{\tensor{\mathring{\Gamma}}{#1}}%eu
\newcommand{\contorsion}[1]{\tensor{K}{#1}}%eu
\newcommand{\n}[1]{\tensor{N}{#1}}%eu
\newcommand{\s}[1]{\tensor{S}{#1}}%eu
\newcommand{\call}{{\cal L}}%eu
\newcommand{\pd}[1]{\tensor{\partial}{#1}}%eu
\newcommand{\nablab}{\bm{\nabla}}%eu
\newcommand{\T}{\bm{T}}%eu
\newcommand{\tp}[1]{\tensor{T}{#1}}%eu teleparallel
\newcommand{\tpt}[1]{\tensor{\tilde{T}}{#1}}%eu teleparallel tilde
\newcommand{\potential}[1]{\tensor{\Sigma}{#1}}%eu teleparallel 
\newcommand{\superpotential}[1]{\tensor{\Xi}{#1} }% caso teleparallel-Weyl
\newcommand{\superpotentialtp}[1]{\tensor{\tilde{\Xi}}{#1} }% caso teleparallel-Weyl tilde
\newcommand{\superpotentialb}{\bm{\Xi} }% caso teleparallel-Weyl
\newcommand{\torsion}[1]{\tensor{{\cal T}}{#1} }% caso teleparallel-Weyl
\newcommand{\torsiontp}[1]{\tensor{\tilde{\cal T}}{#1}}% caso teleparallel-Weyl tilde
\newcommand{\calt}{{\cal T}}%eu teleparalle well
\newcommand{\caltt}{\tilde{\cal T}}%eu teleparalle well tilde
\newcommand{\pconnection}[1]{\tensor[^p]{\Gamma}{#1}}%eu
\newcommand{\pk}[1]{\tensor[^p]{K}{#1}}%eu
\newcommand{\rb}{\bar{r}}%eu
\newcommand{\braket}[1]{\left<#1\right>}%eu
\newcommand{\omegabar}[1]{\tensor{\bar{\omega}}{#1}}%eu 
\newcommand{\ebar}[1]{\tensor{\bar{e}}{#1}}%eu
\newcommand{\thetabar}{\bar{\theta}}
\newcommand{\superpotentialbar}[1]{\tensor{\bar{\Xi}}{#1} }% caso teleparallel-Weyl bar
\newcommand{\torsionbar}[1]{\tensor{\bar{\cal T}}{#1}}

\begin{document}

%\preprint{}

\title{Conformal Teleparallel Theories and Weyl Geometry}


\author{J. B. Formiga}
\email[]{jansen@fisica.ufpb.br}
%\thanks{}
\affiliation{Departamento de Física, Universidade Federal da Paraíba, Caixa Postal 5008, 58051-970 João Pessoa, Pb, Brazil}

%\author{}
%\email{}
%\affiliation{}
\date{\today}


\date{\today}

\begin{abstract}
Despite the fact that General Relativity (GR) has been very successful, many alternative theories of gravity have attracted the attention of a significant number of theoretical physicists. Among these theories, we have theories with conformal symmetry. Here, the use of Weyl geometry to deal with conformal teleparallel gravity is reviewed in great detail. As an application, a model that can be set to be equivalent to the Teleparallel Equivalent of General Relativity (TEGR) and is invariant under diffeomorphisms, local Lorentz transformations (LLT) and Weyl transformations (WT) is created. Some $pp$-wave, spherically symmetric and cosmological solutions are obtained. It turns out that the class of possibles solutions is wider than that of TEGR. In addition, the total and the gravitational energies of the universe are calculated and analyzed. 
\end{abstract}


\maketitle

\section{Introduction}
General relativity  is the standard theory of gravity and has passed all experimental tests so far, including the most recent ones regarding the prediction of gravitational waves \cite{PhysRevLett.116.061102,PhysRevLett.116.221101,PhysRevLett.119.161101}. However, there are many open
questions about its behavior in extreme situations and at the quantum level \cite{Faraoni_2011}. Another issue that arises in this theory is the definition of an energy-momentum tensor (EMT) for the gravitational field \cite{ANDP:ANDP201200272}. As a result, we have seen an increasing interest in alternative theories of gravity in the last few decades. Since the geometrical background of GR is the Riemannian geometry, one of the approaches to formulate new theories of gravity is to change the geometry of the spacetime \cite{HEHL19951}. An example is the 
so-called teleparallel theories, whose geometry is the Weitzenböck one. The most famous teleparallel theory, known as Teleparallel Equivalent of General Relativity (TEGR), is  equivalent to GR at the level of the field equations.

In teleparallel theories, it is possible to define the energy-momentum tensor of gravity (EMTG) in a very satisfactory way \cite{ANDP:ANDP201200272}, although it depends on the tetrad field; this tensor is even compatible with the one expected for linearized gravitational waves \cite{doi:10.1002/andp.201800320}. Furthermore, the gravitational field is described by the torsion tensor, the fundamental quantity is the tetrad field $\e{^a_\mu}$, and the components of the affine connection in the tetrad basis reflect the inertial properties of the frame \cite{0264-9381-34-14-145013}, at least in most cases. It is argued that this allows for the separation of inertia and gravitation, which favors the quantization of gravity via teleparallelism \cite{aldrovandi2012teleparallel}.

Most theories of gravity are invariant only under diffeomorphisms and local $SO(3,1)$ transformations. Nonetheless, there is an interest to add also a conformal invariance, since it is believed that this symmetry was important at early stages of the universe or on the small scales \cite{0034-4885-65-5-201,RevModPhys.34.442}. By adding this symmetry to the theory, we may be able to change the behavior of the gravitational field on small scales so that the modified gravitational theory yields a renormalizable and unitary quantum theory of gravity. On the other hand, on large scales, the modified theory is expected to solve the dark matter and dark energy problems. In the case of teleparallelism, many articles on conformal theories have been published in the last few years \cite{ANDP:ANDP201200037,Momeni2014,Silva2016,0264-9381-34-11-115012,doi:10.1142/S0217732317501139}. In some cases, one adds a scalar field that has no relation with the geometry, which requires some ad hoc assumptions on the way the field changes and on its covariant derivative.  Since Weyl geometry has been used to deal with conformal invariance in some alternative theories of gravity such as, for example, Brans-Dicke theory \cite{PhysRevD.89.064047}, one should find natural that this geometry would also go hand in hand with this symmetry in teleparallel theories; and, in fact, it does \cite{PhysRevD.87.067702}. In this paper, the role that Weyl geometry can play in conformal teleparallel theories (CTT)  as well as nonconformal ones in the presence of  a scalar field is reviewed and discussed. It turns out that the CTTs possess an ambiguity with respect to the frame (tetrad ``plus'' scalar field) where the boundary conditions are imposed.  A conformal teleparallel model that can be set to be equivalent to the TEGR when the ``right'' boundary conditions are chosen is constructed in an integrable Weyl geometry, and its Weitzenböckian counterpart is  exhibited. Wave, spherically symmetric, and cosmological solutions for boundary conditions that do not necessarily assure the equivalence with the TEGR are obtained, and their meaning is discussed. Some results are compared to the ones in Ref.~\cite{doi:10.1142/S0217732317501139}, where basically the same model is used (the main differences are the coupling prescription and the definition of the EMTG).

The notation and conventions used in this paper are presented in Sec.~\ref{14122017a}, while a brief review of non-Riemannian geometries is given in Sec.~\ref{14122017b}. A brief review of the theory that mixes  the ideas of teleparallelism and  Weyl geometry, first presented in Ref. \cite{PhysRevD.87.067702}, is given in Sec.~\ref{14122017c}.  Section \ref{08122017b} is devoted to a particular conformal model and its properties. In Sec.~\ref{25082018c} the matter coupling is presented, while the covariant definition (under WT) of the EMTG is given in Sec.~\ref{17122018a}. The $pp$-wave and the spherical solutions are obtained in Sec.~\ref{17122018b}, while the total and the gravitational  energy of the universe are calculated in Sec.~\ref{17122018c}; a solution is given at the end of this section. Section \ref{14122017d} is dedicated to a final discussion.

\section{Notation and Conventions}\label{14122017a}
The spacetime metric is denoted by $g$ and its signature is $(+,-,-,-)$. In a coordinate basis we have $g_{\mu\nu}$, while in a tetrad basis $\theta^a$ (the frame vectors are $e_a$) we use $\eta_{ab}$; the coframe and the frame satisfy the relation $\theta^a(e_b)=\delta^a_b$. It is clear that Greek letters represent spacetime indices while Latin letters represent tangent space ones, except for Latin letters in the middle of the alphabet ($i,j,k,\ldots$), which stand for spatial coordinate indices. The components of the frame $e_a$ and the coframe $\theta^a$ in a coordinate basis are represented by $\tensor{e}{_a^\mu}$ and $\tensor{e}{^a_\mu}$, respectively. To distinguish indices when numbering, the tangent space indices are used between parentheses: $\tensor{e}{_{(0)}^0}$, $\tensor{e}{_{(1)}^2}$ etc.

Let $V,U,W$ be vector fields. In a spacetime with an affine connection $\nablab$, the curvature tensor can be defined as
\begin{equation}
\bm{R}(V,U)W\equiv\nablab_V\nablab_U W-\nablab_U\nablab_V W-\nablab_{[V,U]}W, \label{07112017b}
\end{equation}
where $[V,U]$ is the Lie bracket of $V$ and $U$. The components of $\bm{R}(V,U)W$ in a coordinate basis are given by
\begin{equation}
\tensor{R}{^\alpha _\mu_\beta_\nu}=\partial_{\beta}\connection{^\alpha_\nu_\mu}-\partial_{\nu}\connection{^\alpha_\beta_\mu}+\connection{^\phi_\nu_\mu}\connection{^\alpha_\beta_\phi}-\connection{^\phi_\beta_\mu}\connection{^\alpha_\nu_\phi}, \label{07112017c}
\end{equation}
where\footnote{The symbol $\braket{df,V}$ is used to represent the action of a $1$-form $df$ on a vector field $V$, that is, $\braket{df,V}=V[f]=V^{\mu}\pd{_\mu}f$, where $V^{\mu}$ are the components of $V$ in the basis $\pd{_\mu}$. For more details on this notation, see Ref.~\cite{Nakahara}. } $\tensor{R}{^\alpha _\mu_\beta_\nu}\equiv\braket{dx^{\alpha},\bm{R}(\pd{_\beta},\pd{_\nu})\pd{_\mu}}$, and $\connection{^\lambda_\mu_\nu}$ are the components of the affine connection $\nablab$ in a coordinate basis, which has been defined as $\connection{^\lambda_\mu_\nu}\equiv \left<dx^{\lambda},\nablab_{\mu}\partial_{\nu} \right>$; notice that, here, the order of all indices matters because of the torsion contributions. In turn, the components of $\nablab$ in the tetrad basis are denoted by  $\tensor{\omega}{^a_b_c}\equiv \left<\theta^a,\nablab_b e_c \right>$. The components of the covariant derivative of a tensor $V$ are defined as $\nabla_{\nu}V^{\mu}\equiv\left<dx^{\mu},\nablab_{\nu}V\right>$. When denoting components of a covariant derivative that also ``acts'' on tetrad indices, the letter $D$ is used. Examples: $D_{\nu}\tensor{e}{_a^\mu}=\partial_{\nu}\tensor{e}{_a^\mu}+\connection{^\mu_\nu_\lambda}\tensor{e}{_a^\lambda}-\tensor{\omega}{^b_\nu _a}\tensor{e}{_b^\mu}$, but, on the other hand $\nabla_{\nu}\tensor{e}{_a^\mu}=\partial_{\nu}\tensor{e}{_a^\mu}+\connection{^\mu_\nu_\lambda}\tensor{e}{_a^\lambda}$.

The torsion tensor and its components in the basis $e_a$ are defined as
\begin{equation}
\T(V,U)\equiv\nablab_V U-\nablab_U V- [V,U], \label{07112017d}
\end{equation}
$\torsion{^a_b_c}\equiv\left<\theta^a,\T(e_b,e_c)\right>$, respectively. From these components, we may define $\torsion{_a}\equiv\torsion{^b_b_a}$.

 From the metric $g$ and the connection $\nablab$, we can define the nonmetricity tensor $Q(V,U,W)$ through $Q(V,U,W)\equiv\left(\nablab_{W}g\right)(V,U)$, whose components are $Q_{\mu\nu\lambda}\equiv\nabla_{\lambda}g_{\mu\nu}$.

The antisymmetric part of a tensor is represented by $A_{[ab]}\equiv(A_{ab}-A_{ba})/2$, while the symmetric one is $A_{(ab)}\equiv(A_{ab}+A_{ba})/2$

\section{Non-Riemannian Geometries}\label{14122017b}
There are many different types of non-Riemannian geometries \cite{0034-4885-65-5-201,HEHL19951,PUETZFELD200559,doi:10.1142/S0219887808002898}. However, here we will deal only with those that generalize Riemannian geometry by relaxing the assumptions that both torsion and nonmetricity tensors vanish. In this case, the affine connection can be written in the form
\begin{equation}
\connection{^\lambda_\mu_\nu}=\chr{^\lambda_\mu_\nu}+\contorsion{^\lambda_\mu_\nu}+\n{^\lambda_\mu_\nu}, \label{17112017a}
\end{equation}
where 
\begin{equation}
\n{^\lambda_\mu_\nu}\equiv-\frac{1}{2}\left(\tensor{Q}{^\lambda_\mu_\nu}+\tensor{Q}{^\lambda_\nu_\mu}-\tensor{Q}{_\mu_\nu^\lambda} \right), \label{17112017b}
\end{equation}
$\contorsion{^\alpha_\nu_\mu}$ is the contorsion tensor, given by
\begin{equation}
\contorsion{^\lambda_\mu_\nu}\equiv-\frac{1}{2}\left(\torsion{_\mu_\nu^\lambda}+\torsion{_\nu_\mu^\lambda}-\torsion{^\lambda_\mu_\nu} \right),\label{17112017c}
\end{equation}
and $\chr{^\alpha_\nu_\mu}$ are the Christoffel symbols. On the other hand, if we use a tetrad basis, the components of the affine connection $\nablab$, denoted by $\tensor{\omega}{^a_b_c}$, will take the form
\begin{equation}
\tensor{\omega}{^a_b_c}=\frac{1}{2}\left(\tensor{\Omega}{_b_c^a}+\tensor{\Omega}{_c_b^a}-\tensor{\Omega}{^a_b_c} \right)+\tensor{K}{^a_b_c}+\tensor{N}{^a_b_c},
\end{equation} 
where $\tensor{\Omega}{^a_b_c}\equiv-\left<\theta^a,[e_b,e_c]\right>$ is the object of anholonomity. These components are frequently called ``spin connection''.

Using $\s{^\lambda_\mu_\nu}=\contorsion{^\lambda_\mu_\nu}+\n{^\lambda_\mu_\nu}$, one can verify that
\begin{eqnarray}
\tensor{R}{^\alpha_\mu_\beta_\nu}&=&\R{^\alpha_\mu_\beta_\nu}+\nabla_{\beta}\s{^\alpha_\nu_\mu}-\nabla_{\nu}\s{^\alpha_\beta_\mu}+\s{^\alpha _\nu_\phi}\s{^\phi _\beta_\mu}
\nonumber\\
&&-\s{^\alpha _\beta_\phi}\s{^\phi _\nu_\mu}+\torsion{^\phi_\beta_\nu}\s{^\alpha_\phi_\mu},
\end{eqnarray}
where $\R{^\alpha_\mu_\beta_\nu}$ is the Riemannian tensor written in terms of the Christoffel symbols.

It follows that, if we take $\tensor{R}{_\mu_\nu}=\tensor{R}{^\alpha_\mu_\alpha_\nu}$, then
\begin{eqnarray}
\tensor{R}{_\mu_\nu}&=&\R{_\mu_\nu}+\nabla_{\alpha}\s{^\alpha_\nu_\mu}-\nabla_{\nu}\s{^\alpha_\alpha_\mu}+\s{^\alpha _\nu_\phi}\s{^\phi _\alpha_\mu}
\nonumber\\
&&-\s{^\alpha _\alpha_\phi}\s{^\phi _\nu_\mu}+\torsion{^\phi_\alpha_\nu}\s{^\alpha_\phi_\mu}.
\end{eqnarray}
Finally, contracting with $g^{\mu\nu}$, we obtain
\begin{eqnarray}
R&=&\R{}+g^{\mu\nu}\nabla_{\alpha}\s{^\alpha_\nu_\mu}-g^{\mu\nu}\nabla_{\nu}\s{^\alpha_\alpha_\mu}+\s{^\alpha ^\mu_\phi}\s{^\phi _\alpha_\mu}
\nonumber\\
&&-\s{^\alpha _\alpha_\phi}\s{^\phi ^\mu_\mu}+\torsion{^\phi_\alpha^\mu}\s{^\alpha_\phi_\mu}.
\end{eqnarray}
In terms of the Riemannian covariant derivative, these identities become
\begin{eqnarray}
\tensor{R}{^\alpha _\mu_\beta_\nu}&=&\R{^\alpha _\mu_\beta_\nu}+\rcd{_\beta}\tensor{S}{^\alpha_\nu_\mu}-\rcd{_\nu}\tensor{S}{^\alpha_\beta_\mu}
+\tensor{S}{^\alpha_\beta_\lambda}\tensor{S}{^\lambda_\nu_\mu}\nonumber\\
&&-\tensor{S}{^\alpha_\nu_\lambda}\tensor{S}{^\lambda_\beta_\mu},
\end{eqnarray}
\begin{eqnarray}
R_{\mu\nu}&=&\R{_\mu_\nu}+\rcd{_\alpha}\tensor{S}{^\alpha_\nu_\mu}-\rcd{_\nu}\tensor{S}{^\alpha_\alpha_\mu}+\tensor{S}{^\alpha_\alpha_\lambda}\tensor{S}{^\lambda_\nu_\mu}
\nonumber\\
&&-\tensor{S}{^\alpha_\nu_\lambda}\tensor{S}{^\lambda_\alpha_\mu},
\end{eqnarray}
\begin{eqnarray}
R&=&\R{}+\rcd{_\alpha}\tensor{S}{^\alpha^\mu_\mu}-\rcd{_\mu}\tensor{S}{^\alpha_\alpha^\mu}+\tensor{S}{^\alpha_\alpha_\lambda}\tensor{S}{^\lambda^\mu_\mu}
\nonumber\\
&&-\tensor{S}{^\alpha^\mu_\lambda}\tensor{S}{^\lambda_\alpha_\mu}.
\end{eqnarray}

\subsection{Teleparallel theories}
A teleparallel theory is a theory based upon a special case of the Riemann-Cartan geometry known as Weitzenböck geometry. In this kind of theory it is assumed that the spacetime curvature vanishes while the torsion tensor plays the role of the gravitational field. One also assumes that the connection is compatible with the metric, i.e., the covariant derivative of the metric tensor is zero. Another way of putting it is to say that there exists a particular tetrad basis $\{e_a\}$ and a connection $\nablab$ such that
\begin{equation}
\nablab_{\mu}e_a=0. \label{24122017a}
\end{equation}
 We see from Eq.~(\ref{07112017d}) that, in this basis, the torsion components take on the form
\begin{equation}
\tp{^a_b_c}=\tensor{\Omega}{^a_b_c}=2\tensor{e}{_b^\mu}\tensor{e}{_c^\nu}   \pd{_{[\mu|}}\tensor{e}{^a_{|\nu]}}. \label{05112017a}
\end{equation}

The frame that satisfies Eq.~(\ref{24122017a}) is a special one, since the components of the affine connection vanish. In a general frame $\ebar{_a}$, related to $e_b$ via  $\ebar{_a}=\tensor{\Lambda}{^b_a}e_b $, where $(\tensor{\Lambda}{^b_a})$ is a Lorentz matrix, Eqs.~(\ref{24122017a})-(\ref{05112017a}) can be recast as $\nablab_{\mu}\ebar{_a}=\tensor{\bar{\omega}}{^b_\mu _a}\ebar{_b}=\tensor{\Lambda}{_c^b}\left(\pd{_\mu}\tensor{\Lambda}{^c_a}\right)\ebar{_b}$ and $\tensor{\bar{T}}{^a_b_c}=2\tensor{\Lambda}{_d^a}\ebar{_{[b|}^\mu} \pd{_\mu} \tensor{\Lambda}{^d_{|c]}} + 2\ebar{_b^\mu}\ebar{_c^\nu}   \pd{_{[\mu|}}\ebar{^a_{|\nu]}}$. The reader should keep in mind that the theory that has been presented in this section is a teleparallel theory with an inertial spin connection. Every time an overbar is used in a component of an object it is because this component is written in a frame where the affine connection does not necessarily vanish. Nonetheless, in the frame $e_a$, it will always vanish (except after the introduction of the Weyl field). 

Teleparallel theories say nothing about the tetrad that one should use in Eq.~(\ref{24122017a}). However, when interpreting quantities that depend on $\e{_a}$, such as the EMTG, we should avoid taking tetrad fields that are anholonomic even in Minkowski spacetime as the frame $e_a$, because these quantities can be affected by fictitious contributions. To see this, let us take the frame $\e{_a}=(\pd{_t},\pd{_r},\frac{1}{r}\pd{_\theta},\frac{1}{r\sin\theta}\pd{_\phi})$, whose coframe is $\theta^a=(dt,dr,rd\theta,r\sin\theta d\phi)$, where ($r$,$\theta$,$\phi$) is a spherical coordinate system. By reading the components of the coframe, $\e{^a_\mu}=diag(1,1,r,r\sin\theta)$, we clearly see that $g_{\mu\nu}=diag(1,-1,-r^2,-r^2\sin^2\theta)$ everywhere (Minkowski spacetime). From Eq.~(\ref{07112017d}), we see that if we choose this frame to satisfy $\nabla_{\mu}e_a=0$, then $\T(e_b,e_c)=\nablab_b e_c-\nablab_c e_b- [e_b,e_c]=- [e_b,e_c]=\tensor{\Omega}{^a_b_c}e_a$, which leads to the following nonvanishing torsion in Minkowski spacetime: $\tp{^{(2)}_{(1)(2)}}=\tp{^{(3)}_{(1)(3)}}=1/r$, and $ \tp{^{(3)}_{(2)(3)}}=\cos\theta/(r\sin\theta)$. It is clear in this example that, by choosing a frame $e_a$ that is not holonomic in Minkowski spacetime simultaneously with the condition $\nabla_{\mu}e_a=0$, the torsion tensor becomes meaningless. Of course, we can still work with a frame $\ebar{_a}$, given by  $\ebar{_a}=\tensor{\Lambda}{^b_a}e_b $, that is not holonomic in Minkowski spacetime.

If, instead of $\theta^a=(dt,dr,rd\theta,r\sin\theta d\phi)$, we had chosen the Cartesian tetrad $\theta^a=(dt,dx,dy,dz)$ to be the one that satisfies $\nabla_{\mu}e_a=0$, the torsion tensor would vanish. It is clear that torsion is not invariant under the choice of the frame that satisfies this condition: it is only invariant under a local Lorentz transformation. Furthermore, quantities defined solely by combinations of the torsion tensor, such as the superpotential, will have the same feature.  Nonetheless, the teleparallel model of the next subsection has field equations that do not depend on the choice of the tetrad field that satisfies Eq.~(\ref{24122017a}), which means that we can use any frame $e_a$ associated with a given spacetime metric $g$ to solve the field equations. 

The aforementioned problem with the choice of $e_a$ does not seem to be well known in the literature. Nevertheless, it has already been addressed in a different way in subsections IV.A and IV.B of Ref.~\cite{PhysRevD.80.064043}.

\subsubsection{Teleparallel equivalent of general relativity}
A particular case of a teleparallel theory is the so called Teleparallel Equivalent of General Relativity (TEGR), whose main feature is to be equivalent to General Relativity (GR) at the level of the field equations. Despite being equivalent to GR it is conceptually different, one example being the possibility of defining an EMTG (see Ref. \cite{ANDP:ANDP201200272} and references therein). 

To construct the TEGR, one uses the identity
%arquivo de 2017/11/02
\begin{equation}
%R=\R{}+\Sigma^{abc}T_{abc}-2\rcd{_\mu}\tp{^\mu}=0
R=\R{}+\tp{}-2\rcd{_\mu}\tp{^\mu}=0,
\end{equation}
where $\rcd{_\mu}$ is the Riemannian covariant derivative, and
\begin{equation}
\tp{}=\frac{1}{4}\tp{^a^b^c}\tp{_a_b_c}+\frac{1}{2}\tp{^a^b^c}\tp{_b_a_c}-\tp{^a}\tp{_a}. \label{14072017e}
\end{equation}
This identity is used to recast the Einstein-Hilbert Lagrangian density in the form ${\cal L}=e\R{}=-e\tp{}+2e\rcd{_\mu}\tp{^\mu}$, where\footnote{The identity $\det(\tensor{e}{^a_\mu})=+\sqrt{-\det(g)}$ holds only when the tetrad is chosen in a particular order. There is no loss of generality in choosing a tetrad field whose determinant is positive.    } $e=\det(\tensor{e}{^a_\mu})=\sqrt{-\det(g)}$. Since at the level of the action the total divergence term is integrated out,  the Lagrangian density ${\cal L}=-e\tp{}$ yields the same field equations as the Einstein-Hilbert one. Thus, in the TEGR, one only focuses  on ${\cal L}=-e\tp{}$.

In dealing with the TEGR and also with $f(\tp{})$ theories, it is convenient to define an object $\potential{^a^b^c}$ (sometimes called superpotential) through
\begin{equation}
\potential{^a^b^c}=\frac{1}{4}\left( \tp{^a^b^c}+\tp{^b^a^c}-\tp{^c^a^b}\right)+\frac{1}{2}\left(\eta^{ac}\tp{^b}-\eta^{ab}\tp{^c} \right). \label{10112017l}
\end{equation}
Note that $\potential{^a^b^c}=-\potential{^a^c^b}$. One can easily show that 
\begin{equation}
\tp{}=\potential{^a^b^c}\tp{_a_b_c}. \label{10112017a}
\end{equation}

The field equations of the TEGR are \cite{ANDP:ANDP200510161}
\begin{equation}
e\tensor{e}{_c^\lambda}\tp{}-4e\potential{^a^b^\lambda}\tp{_a_b_c}+4\partial_{\nu}\left(e\potential{_c^\lambda^\nu}\right)=Ce\tensor{T}{_c^\lambda}, \label{11122017a}
\end{equation}
where $C$ is a constant and $\tensor{T}{_c^\lambda}$ is the energy-momentum tensor of the matter field.

The EMTG in the TEGR is defined as 
\begin{equation}
t^{\lambda\mu}=\frac{c^4}{16\pi G}\left(4\potential{^b^c^\lambda}\tp{_b_c^\mu}-g^{\lambda\mu}\tp{} \right), \label{22072018c}
\end{equation}
where $c$ is the speed of light and $G$ is the gravitational constant. This tensor predicts very interesting and satisfactory results \cite{ANDP:ANDP201200272}, including the energy-momentum tensor of gravitational waves \cite{doi:10.1002/andp.201800320}.

\subsection{Weyl geometry}
Weyl geometry is characterized by a torsionless connection but with a nonmetricity $Q(V,U,W)=\left(\nablab_Wg\right)(V,U)=\sigma(W)g(V,U)$ \cite{adler1965introduction,blagojevic2001gravitation}, where $\sigma$ is called the Weyl $1$-form. In formulating a theory in this geometry, one may or may not choose to impose a symmetry under Weyl transformations (WT), which, in terms of a tetrad field, are given by
\begin{equation}
\tilde{\theta}^a=e^{\theta}\theta^a, \label{05072017b}
\end{equation}
\begin{equation}
\tilde{\sigma}=\sigma+2d\theta. \label{05072017c}
\end{equation}
When $\sigma$ is an exact $1$-form (say, $\sigma=d\varphi$) the transformation (\ref{05072017c}) can be expressed
as
\begin{equation}
\tilde{\varphi}=\varphi+2\theta. \label{10112017c}
\end{equation}

From the fact that the Minkowski metric $\eta_{ab}$ does not change and $g\equiv ds^2=\eta_{ab}\theta^a \theta^b$, we have
\begin{equation}
\tensor{\tilde{e}}{^a_\mu}=e^{\theta}\tensor{e}{^a_\mu},\ \tensor{\tilde{e}}{_a^\mu}=e^{-\theta}\tensor{e}{_a^\mu},\ \tilde{e}=e^{4\theta}e,\ \tilde{g}=e^{2\theta}g. \label{09112017a}
\end{equation} 

In what follows, we present a theory that mixes torsion with the nonmetricity of Weyl geometry.

\section{Mixing Weyl Field with Torsion}\label{14122017c}
There are many ways to mix torsion and nonmetricity.  An interesting way to mix the ideas of teleparallelism with Weyl geometry is to assume that there exists a basis $\{e_a\}$ that satisfies 
\begin{equation}
\nablab_{b}e_a=-\frac{1}{2}\sigma_b e_a, \label{07112017a}
\end{equation}
which, in terms of components, gives $\tensor{\omega}{^a_\mu_b}=-(1/2)\sigma_{\mu}\delta^a_b$. To see that Eq.~(\ref{07112017a}) leads to $\nabla_{\lambda}g_{\mu\nu}=\sigma_{\lambda}g_{\mu\nu}$, we just need to realize that this equation implies $\nablab_{b}\theta^a=(1/2)\sigma_b\theta^a$ and then  calculate the covariant derivative of $g=\eta_{ab}\theta^a \theta^b$.

The motivation for the choice (\ref{07112017a}) lies in the possibility of treating teleparallel theories with conformal invariance in an easier and more fundamental way when one takes $\sigma=d\varphi$ \cite{PhysRevD.87.067702}, where $\varphi$ is a scalar field (the Weyl field).

 Note that Eq.~(\ref{07112017a}) is clearly not invariant under a local SO(3,1) transformation. This is so because it is a gauge choice. To be more precise, in an arbitrary $\tensor{\bar{e}}{_a}$, this equation does not hold. Nevertheless, it is invariant under WT. To see this, consider the change of tetrad in Eq.~(\ref{07112017a}) from $e_a$ to $\tilde{e}_a$  as given by Eq.~(\ref{09112017a}):
\begin{eqnarray}
\nablab_{e_b}e_a=e^{\theta}\nablab_{\tilde{e}_b}e^{\theta}\tilde{e}_a&=&e^{\theta}\left(e^{\theta}\tilde{e}_b[\theta]\tilde{e}_a+e^{\theta}\nablab_{\tilde{b}}\tilde{e}_a \right)
\nonumber\\
&=&e^{2\theta}\tilde{e}_b[\theta]\tilde{e}_a+e^{2\theta}\nablab_{\tilde{b}}\tilde{e}_a, \label{09112017b}
\end{eqnarray}
where $\tilde{e}_a[\theta]\equiv\tensor{\tilde{e}}{_a^\mu}\partial_{\mu}\theta$ and $\nablab_{\tilde{b}}\equiv \nablab_{\tilde{e}_b}$. In turn, the right-hand side of Eq.~(\ref{07112017a}) becomes
\begin{eqnarray}
-\frac{1}{2}\sigma_b e_a=-\frac{1}{2}\tensor{e}{_b^\mu}\sigma_{\mu}e_a&=&-\frac{1}{2}e^{\theta}\tensor{\tilde{e}}{_b^\mu}\left(\tilde{\sigma}_{\mu}-2\partial_{\mu}\theta\right)e^{\theta}\tilde{e}_a
\nonumber\\
&=&-\frac{1}{2}e^{2\theta}\tilde{\sigma}_b \tensor{\tilde{e}}{_a}+e^{2\theta}\tilde{e}_b[\theta]\tilde{e}_a, \label{09112017c}
\end{eqnarray}
where Eqs.~(\ref{05072017c}) and (\ref{09112017a}) have been used. By equating Eq.~(\ref{09112017b}) with (\ref{09112017c}), we obtain $\nablab_{\tilde{b}}\tilde{e}_a=-\tilde{\sigma}_b\tilde{e}_a/2$, which shows the covariance of Eq.~(\ref{07112017a}) under WT.

Taking $V=\partial_{\mu}$, $U=\partial_{\nu}$, and $W=e_b$ in Eq.~(\ref{07112017b}) and using (\ref{07112017a}), one easily finds 
\begin{equation}
\tensor{R}{^a_b_\mu_\nu}=\delta^a_b\partial_{[\nu}\sigma_{\mu]},
\end{equation}
where $\tensor{R}{^a_b_\mu_\nu}=\braket{\theta^a,\bm{R}(\partial_{\mu},\partial_{\nu})e_b}$. In turn, Eq.~(\ref{07112017d}) yields
\begin{equation}
\torsion{^a_b_c}=\tp{^a_b_c} +\tensor{\sigma}{_{[c|}}\delta^a_{|b]}, \label{29102017a}
\end{equation}
where $\tp{^a_b_c}$ is given by Eq.~(\ref{05112017a}).

In this geometry, the quadratic terms $\torsion{^c}\torsion{_c}$, $\torsion{^a^b^c}\torsion{_a_b_c}$, and $\torsion{^a^b^c}\torsion{_b_a_c}$ written in the basis that satisfies (\ref{07112017a}) become
\begin{eqnarray}
\torsion{^c}\torsion{_c}&=&\tp{^c}\tp{_c}+3\sigma^c\tp{_c}+\frac{9}{4}\sigma_a\sigma^a, \label{26102018a}
\end{eqnarray}
\begin{eqnarray}
\torsion{^a^b^c}\torsion{_a_b_c}&=&\tp{^a^b^c}\tp{_a_b_c}+2\sigma^c\tp{_c}+\frac{3}{2}\sigma_a\sigma^a, \label{26102018b}
\end{eqnarray}
\begin{eqnarray}
\torsion{^a^b^c}\torsion{_b_a_c}&=&\tp{^a^b^c}\tp{_b_a_c}+\sigma^c\tp{_c}+\frac{3}{4}\sigma_a\sigma^a. \label{26102018c}
\end{eqnarray}

Let us now prove that, under the transformations (\ref{05072017b})-(\ref{05072017c}), we have $\torsion{^a_b_c}=e^{\theta}\torsiontp{^a_b_c}$. Using (\ref{09112017a}) in the first term on the right-hand side of Eq.~(\ref{29102017a}) [see also Eq.~(\ref{05112017a})], we obtain
\begin{align}
\torsion{^a_b_c}&=e^{2\theta}\tensor{\tilde{e}}{_b^\mu}\tensor{\tilde{e}}{_c^\nu}\left[\pd{_\mu}\left(e^{-\theta}\tensor{\tilde{e}}{^a_\nu}\right) -\pd{_\nu}\left(e^{-\theta}\tensor{\tilde{e}}{^a_\mu}\right) \right]+\tensor{\sigma}{_{[c|}}\delta^a_{|b]}
\nonumber\\
&=2e^{\theta}\tensor{\tilde{e}}{_b^\mu}\tensor{\tilde{e}}{_c^\nu}   \pd{_{[\mu|}}\tensor{\tilde{e}}{^a_{|\nu]}}-2e^{\theta}\tensor{\tilde{e}}{_{[b|}^\mu}\delta^a_{|c]}\pd{_\mu}\theta+\tensor{\sigma}{_{[c|}}\delta^a_{|b]}. \label{25082018a}
\end{align}
From Eq.~(\ref{05072017c}), we see that $\sigma_{\mu}=\tilde{\sigma}_{\mu}-2\pd{_\mu}\theta$, which multiplied by $\tensor{\tilde{e}}{_a^\mu}$ yields 
\begin{equation}
\sigma_a=e^{\theta}\tilde{\sigma}_a-2e^{\theta}\tensor{\tilde{e}}{_a^\mu}\pd{_\mu}\theta, \label{25082018b}
\end{equation}
where $\sigma_{\mu}\tensor{\tilde{e}}{_a^\mu}=\sigma_{\mu}e^{-\theta}\tensor{e}{_a^\mu}=e^{-\theta}\sigma_a$ and $\tilde{\sigma}_a=\tilde{\sigma}_{\mu}\tensor{\tilde{e}}{_a^\mu}$ have been used. Now, using Eq.~(\ref{25082018b}) in the third term on the right-hand side of Eq.~(\ref{25082018a}), we find that $\tensor{\sigma}{_{[c|}}\delta^a_{|b]}=e^{\theta}\tilde{\sigma}_{[c|}\delta^a_{|b]}+2e^{\theta}\tensor{\tilde{e}}{_{[b|}^\mu}\delta^a_{|c]}\pd{_\mu}\theta$. Finally, if we substitute this expression into Eq.~(\ref{25082018a}), we will arrive at $\torsion{^a_b_c}=e^{\theta}\torsiontp{^a_b_c}$ with $\torsiontp{^a_b_c}$ given by 
\begin{equation}
\torsiontp{^a_b_c}=\tpt{^a_b_c} +\tensor{\tilde{\sigma}}{_{[c|}}\delta^a_{|b]}, \label{08122017a}
\end{equation}
where $\tpt{^a_b_c}=2\tensor{\tilde{e}}{_b^\mu}\tensor{\tilde{e}}{_c^\nu}   \pd{_{[\mu|}}\tensor{\tilde{e}}{^a_{|\nu]}}$. Since $\eta_{ab}$ does not change, it is straightforward to verify that $\torsiontp{_c}=e^{-\theta}\torsion{_c}$. In short, we have
\begin{equation}
\torsiontp{^a_b_c}=e^{-\theta}\torsion{^a_b_c},\qquad \torsiontp{_c}=e^{-\theta}\torsion{_c} \label{10112017b}
\end{equation}
under the transformations (\ref{05072017b}) and (\ref{05072017c}). 

A quick look at Eq.~(\ref{10112017b}) shows that 
\begin{align}
\torsion{^a^b^c}\torsion{_a_b_c}=e^{2\theta}\torsiontp{^a^b^c}\torsiontp{_a_b_c},\ \torsion{^a^b^c}\torsion{_b_a_c}=e^{2\theta}\torsiontp{^a^b^c}\torsiontp{_b_a_c},
\nonumber\\
 \torsion{^c}\torsion{_c}=e^{2\theta}\torsiontp{^c}\torsiontp{_c}. \label{19122018d}
\end{align}
This property will allow us to construct models that are invariant under WT.

For the sake of generality, consider a generalized superpotential $\superpotential{^a^b^c}$ defined by the expression
\begin{equation}
\superpotential{^a^b^c}=A\torsion{^a^b^c} +\frac{B}{2}\left( \torsion{^b^a^c}-\torsion{^c^a^b}\right)+\frac{C}{2}\left(\eta^{ab}\torsion{^c}-\eta^{ac}\torsion{^b} \right). \label{20072017a}
\end{equation}
Note that when 
\begin{equation}
A=1/4,\ B=1/2,\ C=-1, \label{02112018a}
\end{equation}
the superpotential $\superpotential{^a^b^c}$ assumes a form analogous to that of $\potential{^a^b^c}$ [see Eq.~(\ref{10112017l})], although they would still be different entities. The scalar $\tp{}$ as defined by Eq.~(\ref{14072017e}) can be generalized to
\begin{equation}
\calt=\superpotential{^a^b^c}\torsion{_a_b_c}=A\torsion{^a^b^c}\torsion{_a_b_c}+B\torsion{^a^b^c}\torsion{_b_a_c}+C\torsion{^a}\torsion{_a}, \label{09112017d}
\end{equation}
which, under WT, clearly transforms like [see, e.g., Eq.~(\ref{19122018d})]
\begin{equation}
\tilde{\calt}=e^{-2\theta}\calt, \label{19122018c}
\end{equation} 
regardless of the values of the parameters $A$, $B$, and $C$.

\subsection{Integrable Weyl field}
For an exact $1$-form $\sigma=d\varphi$, we can rewrite Eq.~(\ref{07112017a}) as
\begin{equation}
\nablab_{\mu}e_a=-\frac{1}{2}\left(\pd{_\mu}\varphi \right) e_a. \label{19122018a}
\end{equation}
It is interesting to note that the condition (\ref{19122018a}) can be turned into the teleparallel one $\nablab_{\tilde{b}}\tilde{e}_a=0$ as long as $\tilde{\varphi}=0$ (recall that $\tilde{b}\equiv\tilde{e}_b$). To see that this is true, we can 
change $\varphi$ to $\tilde{\varphi}$ using Eq.~(\ref{10112017c}) and choose $\theta=-\varphi/2$, which implies $\tilde{\varphi}=0$. Therefore, since (\ref{19122018a}) is covariant, we obtain $\nablab_{\tilde{b}}\tilde{e}_a=0$. As a result,  the integrable Weyl field case can always be turned into a teleparallel theory with a scalar field $\varphi$. Because of this feature, the frame with $\tilde{\varphi}=0$ will be called the ``teleparallel frame''.

From Eqs.~(\ref{10112017c})-(\ref{09112017a}), we can summarize the relation between a general Weyl frame ($\e{_a^\mu}$, $\varphi$) and the teleparallel one ($\tensor{\tilde{e}}{_a^\mu}$,  $\tilde{\varphi}=0$) as follows:
\begin{equation}
\begin{array}{ccc}
\theta=-\varphi/2, & \tensor{\tilde{e}}{_a_\mu}=e^{-\varphi/2}\e{_a_\mu},&  \tensor{\tilde{e}}{_a^\mu}=e^{\varphi/2}\e{_a^\mu},
\\
\tilde{g}_{\mu\nu}=e^{-\varphi}g_{\mu\nu}, & \tilde{e}=e^{-2\varphi}e, & \torsiontp{^a_b_c}=e^{\varphi/2}\torsion{^a_b_c}.
\end{array} \label{08112018a}
\end{equation}

Although the frame  ($\e{_a^\mu}$, $\varphi$) has an arbitrary Weyl field $\varphi$, the vector field $\e{_a}$ is still the one that satisfies the gauge (\ref{19122018a}). To avoid confusion, a general frame will be denoted by ($\ebar{_a}$, $\varphi$), where $\ebar{_a}$ does not necessarily satisfy this gauge, but is related to $\e{_a}$ by means of a LLT.

\section{The case $ee^{-\varphi}\calt$ with the parameters (\ref{02112018a})}\label{08122017b}
To construct a model that is invariant under WT and at the same time becomes equivalent to the TEGR when $\varphi$ vanishes, one can work  with the Lagrangian density $\call_g=ee^{-\varphi}\torsion{}$, where $\torsion{}$ corresponds to Eq.~(\ref{09112017d}) with $A=1/4$, $B=1/2$, $C=-1$, and an affine connection $\nablab$ that satisfies Eq.~(\ref{19122018a}). From now on, we will deal only with this model and its properties.

To see that $\call_g$ is invariant under the transformations (\ref{05072017b}) and (\ref{10112017c}), that is $ee^{-\varphi}\calt=\tilde{e}e^{-\tilde{\varphi}}\caltt$, we just need to use the expressions in (\ref{09112017a}) and Eq.~(\ref{19122018c}). It is also clear that if we go to the teleparallel frame ($\theta=-\varphi/2$), the Lagrangian density will become that of the TEGR.

Treating $\tensor{e}{^a_\lambda}$ and $\varphi$ as independent variables and taking variations of the action $S=\int d^4x ee^{-\varphi}\calt$, we obtain
\begin{eqnarray}
\frac{\delta \call_g}{\delta\tensor{e}{^c_\lambda}}&=&e^{-\varphi}\biggl[e\tensor{e}{_c^\lambda}\calt-4e\superpotential{^a^b^\lambda}\torsion{_a_b_c}
\nonumber\\
&&+4\partial_{\nu}\left(e\superpotential{_c^\lambda^\nu}\right)-2e\left(\partial_{\nu}\varphi\right)\superpotential{_c^\lambda^\nu}\biggr], \label{11072017a}
\end{eqnarray}
\begin{equation}
\frac{\delta\call_g}{\delta \varphi}=e^{-\varphi}\left[-e\calt-2e\torsion{^\mu}\partial_{\mu}\varphi+2(\partial_{\mu}e\torsion{^\mu}) \right]. \label{20072017b}
\end{equation}
This second equation is redundant, as will be shown later. Let us first analyze the covariance of these two equations.

\subsection{Weyl invariance and the equivalence with the TEGR}
From Eqs. (\ref{05072017b})-(\ref{09112017a}), (\ref{10112017b}) and (\ref{19122018c}), we see immediately that 
\begin{eqnarray}
e^{-\varphi}e\tensor{e}{_c^\lambda}\calt&=&e^{\theta}e^{-\tilde{\varphi}}\tilde{e}\tensor{\tilde{e}}{_c^\lambda}\caltt, \label{10112017d}
\\
 -4e^{-\varphi}e\tensor{e}{_d^\lambda}\superpotential{^a^b^d}\torsion{_a_b_c}&=&-4e^{\theta}e^{-\tilde{\varphi}}\tilde{e}\tensor{\tilde{e}}{_d^\lambda}\superpotentialtp{^a^b^d}\torsiontp{_a_b_c}, \label{10112017e}
 \\
-ee^{-\varphi}\calt&=&-\tilde{e}e^{-\tilde{\varphi}}\caltt. \label{10112017f}
\end{eqnarray}
One can also verify that $\superpotential{_c^\lambda^\nu}=\tensor{e}{_a^\lambda}\tensor{e}{_b^\nu}\superpotential{_c^a^b}=e^{3\theta}\superpotentialtp{_c^\lambda^\nu}$. With respect to the other terms of Eqs. (\ref{11072017a}), we have:
\begin{align}
4e^{-\varphi}\partial_{\nu}\left(e\superpotential{_c^\lambda^\nu}\right)=4e^{-\tilde{\varphi}+2\theta}\partial_{\nu}\left(e^{-\theta}\tilde{e}\superpotentialtp{_c^\lambda^\nu}\right)=
\nonumber\\
=e^{\theta}\biggl[4e^{-\tilde{\varphi}}\partial_{\nu}\left(\tilde{e}\superpotentialtp{_c^\lambda^\nu}\right)-4(\pd{_\nu}\theta)e^{-\tilde{\varphi}}\tilde{e}\superpotentialtp{_c^\lambda^\nu}\biggr], \label{10112017g}
\end{align}
and
\begin{align}
-2e^{-\varphi}e\left(\partial_{\nu}\varphi\right)\superpotential{_c^\lambda^\nu}=-2e^{-\tilde{\varphi}+2\theta}e^{-\theta}\tilde{e}\left(\partial_{\nu}\tilde{\varphi}-2\partial_{\nu}\theta\right)\superpotentialtp{_c^\lambda^\nu}=
\nonumber\\
=e^{\theta}\biggl[-2e^{-\tilde{\varphi}}\tilde{e}\left(\partial_{\nu}\tilde{\varphi}\right)\superpotentialtp{_c^\lambda^\nu} +4(\pd{_\nu}\theta)e^{-\tilde{\varphi}}\tilde{e}\superpotentialtp{_c^\lambda^\nu}\biggr]. \label{10112017gb}
\end{align}
Comparing Eqs.~(\ref{10112017g}) and (\ref{10112017gb}), we see that the summation of the last two terms of Eq.~(\ref{11072017a}) also changes by a factor of $e^{\theta}$.

A similar procedure leads to
\begin{eqnarray}
e^{-\varphi}\left[-2e\torsion{^\mu}\partial_{\mu}\varphi+2(\partial_{\mu}e\torsion{^\mu})\right]&=&e^{-\tilde{\varphi}}\bigl[-2\tilde{e}\torsiontp{^\mu}\partial_{\mu}\tilde{\varphi}
\nonumber\\
&&+2(\partial_{\mu}\tilde{e}\torsiontp{^\mu})\bigr]. \label{10112017h}
\end{eqnarray}
By substituting Eqs. (\ref{10112017d})-(\ref{10112017h}) into (\ref{11072017a}) and (\ref{20072017b}), we find that
\begin{equation}
\frac{1}{e^{\theta}}\frac{\delta \call_g}{\delta\tensor{e}{^c_\lambda}}=\frac{\delta \tilde{\call}_g}{\delta\tensor{\tilde{e}}{^c_\lambda}},\qquad \frac{\delta \call_g}{\delta\varphi}=\frac{\delta \tilde{\call}_g}{\delta\tilde{\varphi}}.
\end{equation}

Let us now see whether Eq.~(\ref{11072017a}) has any relationship with (\ref{20072017b}). In doing so, we may choose to work with the teleparallel frame $\tilde{e}_a$. In this case $\tilde{\varphi}=0$ and Eqs. (\ref{11072017a})-(\ref{20072017b})  become
\begin{eqnarray}
\frac{1}{e^{\theta}}\frac{\delta \call_g}{\delta\tensor{e}{^c_\lambda}}&=&\tilde{e}\tensor{\tilde{e}}{_c^\lambda}\caltt-4\tilde{e}\superpotentialtp{^a^b^\lambda}\torsiontp{_a_b_c}
+4\partial_{\nu}\left(\tilde{e}\superpotentialtp{_c^\lambda^\nu}\right), \label{10112017i}
\end{eqnarray}
\begin{equation}
\frac{\delta\call_g}{\delta \varphi}=-\tilde{e}\caltt+2(\partial_{\mu}\tilde{e}\torsiontp{^\mu}). \label{10112017j}
\end{equation}
Applying $\tensor{\tilde{e}}{^c_\lambda}$ on both sides of Eq.~(\ref{10112017i}) and using Eq.~(\ref{09112017a}) we have
\begin{eqnarray}
\tensor{e}{^c_\lambda}\frac{\delta \call_g}{\delta\tensor{e}{^c_\lambda}}&=&4\tilde{e}\caltt-4\tilde{e}\caltt+4\tensor{\tilde{e}}{^c_\lambda}\partial_{\nu}\left(\tilde{e}\superpotentialtp{_c^\lambda^\nu}\right)
\nonumber\\
&=&4\tensor{\tilde{e}}{^c_\lambda}\partial_{\nu}\left(\tilde{e}\superpotentialtp{_c^\lambda^\nu}\right),
\end{eqnarray}
where Eq.~(\ref{09112017d}) has been used in the first line. By putting $\tensor{\tilde{e}}{^c_\lambda}$ on the right side of the partial derivative, we see that  the above expression can be written in the alternative form
\begin{eqnarray}
\tensor{e}{^c_\lambda}\frac{\delta \call_g}{\delta\tensor{e}{^c_\lambda}}&=&4\partial_{\nu}\left(\tilde{e}\superpotentialtp{_c^c^\nu}\right)-4\tilde{e}\superpotentialtp{_c^\lambda^\nu}\partial_{\nu}\tensor{\tilde{e}}{^c_\lambda}
\nonumber\\
&=& -4\partial_{\nu}\left( \tilde{e}\torsiontp{^\nu}\right)-4\tilde{e}\superpotentialtp{_c^\lambda^\nu}\pd{_{[\nu|}}\tensor{\tilde{e}}{^c_{|\lambda]}},
\end{eqnarray}
where $\superpotentialtp{_c^c^\nu}=-\torsiontp{^\nu}$ and the property $\superpotential{^a^b^c}=-\superpotential{^a^c^b}$ have been used in the last line. For $\tilde{\varphi}=0$,  Eq.~(\ref{08122017a}) reduces to $\torsiontp{^a_b_c}=2\tensor{\tilde{e}}{_b^\mu}\tensor{\tilde{e}}{_c^\nu}   \pd{_{[\mu|}}\tensor{\tilde{e}}{^a_{|\nu]}}$, which combined with the above expression gives
\begin{eqnarray}
\tensor{e}{^c_\lambda}\frac{\delta \call_g}{\delta\tensor{e}{^c_\lambda}}&=&-4\partial_{\nu}\left( \tilde{e}\torsiontp{^\nu}\right)+2\tilde{e}\superpotentialtp{_a^b^c}\torsiontp{^a_b_c}
\nonumber\\
&=& -4\partial_{\nu}\left( \tilde{e}\torsiontp{^\nu}\right)+2\tilde{e}\caltt
\nonumber\\
&=&-2\frac{\delta\call_g}{\delta \varphi},
\end{eqnarray}
where the last equality comes from Eq.~(\ref{10112017j}). As it is clear in the equation above, Eq.~(\ref{20072017b}) is proportional to the trace of  (\ref{11072017a}). Notice that this result is independent of the choice $\tilde{\varphi}=0$.

When coupling with matter, we can choose the Lagrangian matter density in such a way that the resultant field equations remain covariant and  equivalent to the TEGR.  As will be shown later, it is possible to preserve this symmetry and even so construct a theory that is not equivalent to TEGR.

\subsection{The symmetry $SO(3,1)$}
In the way that Eq.~(\ref{11072017a}) is written, it is not clear whether this field equation is covariant under LLT. It is clear, though, that Eq.~(\ref{20072017b}) is. To see that Eq.~(\ref{11072017a}) can be written in a covariant form under LLT, we need to get rid of the term with the partial derivative.  This can be done by using the covariant derivative of $\superpotentialb=\superpotentialbar{_c^\lambda^\nu}\thetabar^c\otimes\partial_{\lambda}\otimes\partial_{\nu}$. To avoid confusion between the tetrad basis that satisfies Eq.~(\ref{19122018a}) and a general one, we denote the former by  $\e{_a}$  and the latter by $\ebar{_a}$, while all quantities that depend on $\ebar{_a}$ will be denoted with an overbar. So, we have
\begin{equation}
D_{\nu}\superpotentialbar{_c^\lambda^\nu}=\partial_{\nu}\superpotentialbar{_c^\lambda^\nu}-\omegabar{^a_\nu_c}\superpotentialbar{_a^\lambda^\nu}+\connection{^\lambda_\nu_\mu}\superpotentialbar{_c^\mu^\nu}+\connection{^\nu_\nu_\mu}\superpotentialbar{_c^\lambda^\mu}. \label{11122017b}
\end{equation}
It is worthwhile to remember that $D_{\nu}$ is used for the components of the covariant derivative of an object of the type $\bm{A}=\tensor{A}{_a^\mu}\theta^a\otimes\partial_{\mu}$, while $\nabla_{\nu}$ is used when the nature of the object is $A_a=\tensor{A}{_a^\mu}\partial_{\mu}$: the action of the affine connection $\nablab$ on $\bm{A}$ will result in a ``$D$ component'' while the action on $A_a$ will give a ``$\nabla$ component''.

Since the tetrad $\thetabar^a$ is connected to $\theta^a$ by  $\thetabar^a=\tensor{\Lambda}{_b^a}\theta^b$, where $(\tensor{\Lambda}{^a_b})$ is the Lorentz matrix and $\tensor{\Lambda}{_c^b}=\tensor{(\Lambda^{-1})}{^b_c}$ is its inverse, then the affine connection in one basis is related to the other through the expression
\begin{eqnarray}
\tensor{\bar{\omega}}{^a_b_c}&=&\tensor{\Lambda}{_d^a}\tensor{\Lambda}{^f_b}e_f[\tensor{\Lambda}{^d_c}]+\tensor{\Lambda}{_g^a}\tensor{\Lambda}{^f_b}\tensor{\Lambda}{^d_c}\tensor{\omega}{^g_f_d}
\nonumber\\
&=&\tensor{\Lambda}{_d^a}\tensor{\Lambda}{^f_b}\e{_f^\mu}\pd{_\mu}\tensor{\Lambda}{^d_c}-\frac{1}{2}\delta^a_c\tensor{\Lambda}{^f_b}\e{_f^\mu}\pd{_\mu}\varphi, \label{25082017f}
\end{eqnarray}
where $\tensor{\omega}{^g_f_d}=-(1/2)e_f[\varphi]\delta^g_d$ was used in the second line. In turn, the torsion components are obviously given by
\begin{equation}
\torsionbar{^a_b_c}=\omegabar{^a_b_c}-\omegabar{^a_c_b}+\tensor{\bar{\Omega}}{^a_b_c}, \label{20122018d}
\end{equation}
where  $\tensor{\bar{\Omega}}{^a_b_c}=-\braket{\thetabar^a,[\ebar{_b},\ebar{_c}]}=2\ebar{_b^\mu}\ebar{_c^\nu}   \pd{_{[\mu|}}\ebar{^a_{|\nu]}}$. From these expressions one can easily check that $\torsion{^\lambda_\mu_\nu}=\ebar{_a^\lambda}\ebar{^b_\mu}\ebar{^c_\nu}\torsionbar{^a_b_c}$.

Now, to use Eq.(\ref{11122017b}) in (\ref{11072017a}), we need to rewrite Eq.(\ref{11122017b}) in terms of the basis $\e{_a}$, in which case we have $\tensor{\omega}{^a_\nu_c}=-(1/2)\left(\partial_{\nu}\varphi\right)\delta^a_c$ [see, e.g., Eq.~(\ref{19122018a})]. By using Eqs. (\ref{17112017a})-(\ref{17112017c}), the nonmetricity tensor $Q_{\mu\nu\lambda}=\left(\partial_{\lambda}\varphi\right)g_{\mu\nu}$, and $e=\det(\tensor{e}{^a_\mu})$, one can manipulate Eq.(\ref{11122017b}) to get
\begin{align}
\partial_{\nu}\left(e\superpotential{_c^\lambda^\nu} \right)&=e\biggl[D_{\nu}\superpotential{_c^\lambda^\nu}-\torsion{_\nu}\superpotential{_c^\lambda^\nu}+\frac{3}{2}\left(\partial_{\nu}\varphi\right)\superpotential{_c^\lambda^\nu}
+\frac{1}{2}\torsion{^\lambda_\mu_\nu}\superpotential{_c^\mu^\nu} \biggr], \label{28122018a}
\end{align}
where the identities $\pd{_\nu}e=e\chr{^\mu_\nu_\mu}$, $\tensor{N}{^\nu_\nu_\mu}=-2\pd{_\mu}\varphi$, $\connection{^\lambda_{[\mu\nu]}}=\tensor{K}{^\lambda_{[\mu\nu]}}=(1/2)\torsion{^\lambda_\mu_\nu}$, $\tensor{K}{^\nu_\nu_\mu}=\torsion{_\mu}$ have been used. Notice that Eq.~(\ref{28122018a}) holds only in the basis $\e{_a}$. Finally, substituting Eq.~(\ref{28122018a}) into Eq.~(\ref{11072017a}) gives
\begin{align}
\frac{\delta\call_g}{\delta \tensor{e}{^c_\lambda}}&=ee^{-\varphi}\biggl[\tensor{e}{_c^\lambda}\calt-4\superpotential{^a^b^\lambda}\torsion{_a_b_c}+4D_{\nu}\superpotential{_c^\lambda^\nu}
+4(\partial_{\nu}\varphi-\torsion{_\nu})\superpotential{_c^\lambda^\nu}+2\torsion{^\lambda_\mu_\nu}\superpotential{_c^\mu^\nu} \biggr], \label{11122017c}
\end{align}
which is manifestly covariant, despite being written in terms of the particular basis $\e{_a}$. Using Eq.~(\ref{11122017b}), one can easily check that Eq.~(\ref{11122017c}) reduces\footnote{Note that the letter $D$ in Ref.~\cite{ANDP:ANDP200510161} has a slightly different meaning.} to Eq.~(11) of Ref.~\cite{ANDP:ANDP200510161} for $\varphi=0$.

Since Eq.~(\ref{11122017c}) is covariant, in a general basis $\ebar{_a}$, it can be recast as
\begin{align}
\frac{\delta\call_g}{\delta \ebar{^c_\lambda}}&=ee^{-\varphi}\biggl[\ebar{_c^\lambda}\calt-4\superpotentialbar{^a^b^\lambda}\torsionbar{_a_b_c}+4D_{\nu}\superpotentialbar{_c^\lambda^\nu}
+4(\partial_{\nu}\varphi-\torsion{_\nu})\superpotentialbar{_c^\lambda^\nu}+2\torsion{^\lambda_\mu_\nu}\superpotentialbar{_c^\mu^\nu} \biggr], \label{28122018b}
\end{align}
where $e=\ebar{}$, $\torsion{}=\torsionbar{}$, $\torsion{_\nu}=\torsionbar{_\nu}$, and $\torsion{^\lambda_\mu_\nu}=\torsionbar{^\lambda_\mu_\nu}$.  The relation between Eqs.~(\ref{11122017c}) and (\ref{28122018b}) is $\delta\call_g/\delta \ebar{^c_\lambda}=\tensor{\Lambda}{^b_c}\left(\delta\call_g/\delta \tensor{e}{^b_\lambda}\right)$.

The procedure that we have seen in this section ensures only that, once a tetrad $\e{_a}$ is chosen, a change to a new one ($\ebar{_a}$)  will not alter the form of the field equations. In addition, since these equations are the same as that of the TEGR in the teleparallel frame ($\tilde{\varphi}=0$), they can be written in terms of Einstein's tensor (which depends only on the metric $g_{\mu\nu}$) in this frame. Therefore, given a spacetime $g_{\mu\nu}$, the solution will not depend on the frame we choose to be $\e{_a}$.

Although the field equations are covariant under Weyl and Local Lorentz transformations, the theory does depend on the frame we chose to be the frame $\e{_a}$, i.e., the one that satisfies (\ref{19122018a}) [this also happens with the TEGR when choosing the frame that satisfies Eq.~(\ref{24122017a}) to calculate the EMTG]. In addition, the theory presented here will also depend on which pair ($\e{_a}$,$\varphi$) (we also call this pair ``a frame'') we choose to impose the boundary conditions. For example, if we impose that the spacetime is spherically symmetric and asymptotically flat in the frame ($\e{_a}$, $\tilde{\varphi}=0$), we obtain the Schwarzschild solution and, by performing a WT, we get a set of solutions that are equivalent to the Schwarzschild one. However, if we assume that these boundary conditions hold in a frame with a nontrivial $\varphi$, the solution will not necessarily be equivalent to the Schwarzschild one.  In short, the theory presented here depends on three key points: the field equations, the chosen $\e{_a}$ (only when calculating quantities such as the EMTG), and the frame where the boundary conditions are applied (because of $\varphi$). These properties will become clear in Secs.~\ref{17122018b} and \ref{20122018e}.

\subsection{The teleparallel equivalent of $ee^{-\varphi}\calt$}
From the expressions (\ref{26102018a})-(\ref{26102018c}), we can recast the Lagrangian density $\call_g=ee^{-\varphi}\calt$ as
\begin{align}
\call_p=e\biggl[ \phi^2\left(\frac{1}{4}\tp{^a^b^c}\tp{_a_b_c}+\frac{1}{2}\tp{^a^b^c}\tp{_b_a_c}-\frac{1}{3}\tp{^a}\tp{_a}\right)
-6g^{\mu\nu}\phi_{|\mu}\phi_{|\nu}\biggr], \label{25102018a}
\end{align}
where  $\phi_{|\mu}\equiv (\pd{_\mu}-\tp{_\mu}/3)\phi$ and $\phi=e^{-\varphi/2}$. This is exactly (up to a minus sign) the Lagrangian density of the model with conformal invariance presented in Refs.~\cite{ANDP:ANDP201200037,doi:10.1142/S0217732317501139}. 

In the teleparallel approach, one assumes that under the transformation (\ref{05072017b}) the scalar field $\phi$ transforms like $\tilde{\phi}=e^{-\theta}\phi$ and that its covariant derivative is $\phi_{|\mu}\equiv (\pd{_\mu}-\tp{_\mu}/3)\phi$. By using the relation $\phi=e^{-\varphi/2}$ in the transformation of $\phi$, we arrive at Eq.~(\ref{10112017c}). Thus, the teleparallel model in Refs.~\cite{ANDP:ANDP201200037,doi:10.1142/S0217732317501139} possesses a  hidden Weyl structure. 

Since the independent variables of $\call_p$ in Refs.~\cite{ANDP:ANDP201200037,doi:10.1142/S0217732317501139} and $\call_g$ are the same, the field equations are also the same. It is clear that the model in these references and the one considered here are equivalent, at least in vacuum.

 The connection between the covariant derivative in the approach of Refs.~\cite{ANDP:ANDP201200037,doi:10.1142/S0217732317501139} and the one here can be seen as follows.  We start first with the condition (\ref{24122017a}) in the Weitzenböck geometry then, using only identities, we  write this equation in the form of Eq.~(\ref{07112017a}). In doing so, let us denote the Weitzenböck connection by $\pconnection{^\lambda_\mu_\nu}$. Thus, Eq.~(\ref{24122017a}) can be written as
\begin{equation}
\pd{_\mu}\e{_a^\lambda}+\pconnection{^\lambda_\mu_\nu}\e{_a^\nu}=0, \label{27102018a}
\end{equation}
while the connection $\pconnection{^\lambda_\mu_\nu}$ can be written in the form
\begin{equation}
\pconnection{^\lambda_\mu_\nu}=\chr{^\lambda_\mu_\nu}+\pk{^\lambda_\mu_\nu}, \label{27102018b}
\end{equation} 
where $\chr{^\lambda_\mu_\nu}$ are the Christoffel symbols, and  $\pk{^\lambda_\mu_\nu}$ the Weitzenböck contorsion. On the other hand, the contorsion of the Weyl geometry used here can be expressed in terms of $\pk{^\lambda_\mu_\nu}$ as $\tensor{K}{^\lambda_\mu_\nu}=\pk{^\lambda_\mu_\nu}+\frac{1}{2}\left(\sigma_{\nu}\delta^{\lambda}_{\mu}-\sigma^{\lambda}g_{\mu\nu} \right)$ [see, e.g., Eqs.~(\ref{17112017c}), (\ref{05112017a}), and (\ref{29102017a})].
From Eq.~(\ref{17112017b}) and the fact that we are using $Q_{\lambda\mu\nu}=\sigma_{\nu}g_{\lambda\mu}$, we see that $\tensor{K}{^\lambda_\mu_\nu}+\tensor{N}{^\lambda_\mu_\nu}=\pk{^\lambda_\mu_\nu}-(1/2)\sigma_{\mu}\delta^{\lambda}_{\nu}$, which implies $\pk{^\lambda_\mu_\nu}=\tensor{K}{^\lambda_\mu_\nu}+\tensor{N}{^\lambda_\mu_\nu}+(1/2)\sigma_{\mu}\delta^{\lambda}_{\nu}$. Substituting the latter result into Eq.~(\ref{27102018b}), we find that $\pconnection{^\lambda_\mu_\nu}=\connection{^\lambda_\mu_\nu}+(1/2)\sigma_{\mu}\delta^{\lambda}_{\nu}$, where we have used Eq.~(\ref{17112017a}). Therefore, without loss of generality, Eq.~(\ref{27102018a}) can be rewritten as $\pd{_\mu}\e{_a^\lambda}+\connection{^\lambda_\mu_\nu}\e{_a^\nu}=-(1/2)\sigma_{\mu}\e{_a^\lambda}$, which is exactly the Eq.~(\ref{07112017a}).

The teleparallel frame $\tensor{\tilde{e}}{_a^\lambda}$ corresponds to the case $\tilde{\phi}=1$ ($\tilde{\varphi}=0$). In this frame, $\call_p$ becomes identical to the TEGR Lagrangian density.

\section{Matter Coupling}\label{25082018c}
To investigate the coupling with a matter field, we stick to the model presented in Sec.~\ref{08122017b} and use a procedure very similar to that used in Ref.~\cite{PhysRevD.89.064047}. The action will be taken as
\begin{equation}
S=S_g+S_{\Lambda}+S_M, \label{25082018d}
\end{equation}
where
\begin{equation}
S_g=\int d^4x ee^{-\varphi}\calt \label{25082018e}
\end{equation}
is the geometrical part,
\begin{equation}
S_M=-\frac{8\pi G}{c^4}\int d^4x \call_M \label{25082018f}
\end{equation}
is the matter sector, and
\begin{equation}
S_{\Lambda}=\int d^4x \call_{\Lambda} \label{25082018g}
\end{equation}
is the sector related to the cosmological constant. To make sure that the action is invariant under WT, we must take $\call_{\Lambda}$ as
\begin{equation}
\call_{\Lambda}=-2ee^{-2\varphi}\Lambda, \label{25082018h}
\end{equation}
where $\Lambda$ is the cosmological constant.

In order to obtain the Lagrangian density of a matter field $\psi$, we take $\call_M=ee^{-2\varphi}L_M(\tensor{e}{^c_\lambda},\varphi,\psi)$. The  Lagrangian $L_M(\tensor{e}{^c_\lambda},\varphi,\psi)$ is obtained from the version in GR by changing $\tensor{e}{^c_\lambda}$ to $e^{-\varphi/2}\tensor{e}{^c_\lambda}$ (or $\tensor{e}{_c^\lambda}$ to $e^{\varphi/2}\tensor{e}{_c^\lambda}$), i.e, we take $L_M(\tensor{e}{^c_\lambda},\varphi,\psi)=L_M^{\textrm{\tiny GR}}(e^{-\varphi/2}\tensor{e}{^c_\lambda},\psi)$, where $L_M^{\textrm{\tiny GR}}(\tensor{e}{^c_\lambda},\psi)$ is the matter Lagrangian defined in GR. Notice that, for simplicity, it is assumed that the Lagrangian does not depend on the connection. 

\subsection{Matter energy-momentum tensor}
In general, the energy-momentum tensor of GR is defined as
\begin{equation}
\theta^{\mu\nu}=\frac{1}{e}\frac{\delta (eL_M^{\textrm{\tiny GR}}(\tensor{e}{^c_\lambda},\psi))}{\delta g_{\mu\nu}}.
\end{equation}
Following the procedure defined in the previous section, our definition will be
\begin{equation}
\theta^{\mu\nu}\equiv\frac{1}{ee^{-2\varphi}}\frac{\delta \call_M}{\delta (e^{-\varphi}g_{\mu\nu})},
\end{equation}
where $\call_M=ee^{-2\varphi}L_M^{\textrm{\tiny GR}}(e^{-\varphi/2}\tensor{e}{^c_\lambda},\psi)$. Note that $e\to ee^{-2\varphi}$. This definition ensures that $\theta^{\mu\nu}$ is invariant under WT. However, since we are going to lower indices with $g_{\mu\nu}$, rather than with $e^{-\varphi}g_{\mu\nu}$, any version with different types of indices or index positions may not be invariant under WT ($\tensor{\theta}{_c^\lambda}$, for example).

As we are using the tetrad formalism, it is more convenient to write the definition of the energy-momentum tensor in the equivalent form
\begin{equation}
\tensor{\theta}{_c^\lambda}=\frac{e^{(5/2)\varphi}}{2e}\frac{\delta\call_M}{\delta (e^{-\varphi/2}\tensor{e}{^c_\lambda})}. \label{11112018a}
\end{equation} 

The field equation derived from the action (\ref{25082018d}) is
\begin{equation}
\frac{\delta\left( \call_g+\call_{\Lambda}\right)}{\delta\e{^c_\lambda}}=-2\chi ee^{-3\varphi}\tensor{\theta}{_c^\lambda}, \label{11112018b}
\end{equation}
where, in natural units, $\chi=-8\pi$ (when using $G$ and $c$, we take $\chi=-8\pi G/c^4$).

We do not need to worry about variation with respect to $\varphi$ because the resultant equation is equivalent to the trace of (\ref{11112018b}). To be more precise, we have $\e{^c_\lambda}\delta \call/\delta \e{^c_\lambda}=-2\delta \call/\delta\varphi$, where $\call$ is the total Lagrangian density.

\section{The gravitational energy-momentum tensor}\label{17122018a}
The definition of the EMTG  can be achieved by assuming that the version of the TEGR, Eq.~(\ref{22072018c}),  holds in the teleparallel frame, then we go to a generic Weyl frame. In this case, we must have
\begin{equation}
\tilde{t}^{\lambda\mu}=-\frac{1}{2\chi}\left(4\superpotentialtp{^b^c^\lambda}\torsiontp{_b_c^\mu}-\tilde{g}^{\lambda\mu}\torsiontp{}\right). \label{16112018a}
\end{equation}
From Eqs.~(\ref{19122018c}) and (\ref{08112018a}) we see that $\torsiontp{_b_c^\mu}=e^{\varphi}\torsion{_b_c^\mu}$, $\torsiontp{}=e^{\varphi}\torsion{}$, and $\superpotentialtp{^b^c^\lambda}=e^{\varphi}\superpotential{^b^c^\lambda}$. Therefore, in an arbitrary Weyl frame we have
\begin{equation}
t^{\lambda\mu}=-\frac{e^{2\varphi}}{2\chi}\left(4\superpotential{^b^c^\lambda}\torsion{_b_c^\mu}-g^{\lambda\mu}\torsion{}\right). \label{16112018b}
\end{equation}
With the help of Eq.~(\ref{11072017a}), we can write Eq.~(\ref{11112018b}) in the teleparallel frame as
\begin{equation}
\pd{_\nu}\left(\tilde{e}\superpotentialtp{^a^\lambda^\nu} \right)=-\frac{\chi}{2}\tilde{e}\tensor{\tilde{e}}{^a_\mu}(\tilde{t}^{\lambda\mu}+\tilde{\theta}^{\lambda\mu}-\frac{\Lambda}{\chi}\tilde{g}^{\mu\lambda}). \label{16112018c}
\end{equation}
 Going back to the Weyl frame through the transformations (\ref{08112018a}), we arrive at
\begin{equation}
\pd{_\nu}\left(e^{-\varphi/2}e\superpotential{^a^\lambda^\nu} \right)=-\frac{\chi e^{(-5\varphi/2)}e\e{^a_\mu}}{2}(t^{\lambda\mu}+\theta^{\lambda\mu}-\frac{\Lambda}{\chi}e^{\varphi}g^{\mu\lambda}), \label{16112018d}
\end{equation}
where use of $\tilde{\theta}^{\lambda\mu}=\theta^{\lambda\mu}$, $\tilde{t}^{\lambda\mu}=t^{\lambda\mu}$, and $\superpotentialtp{^a^\lambda^\nu}=e^{3\varphi/2}\superpotential{^a^\lambda^\nu}$ have been made. This equation is equivalent to Eq.~(\ref{11112018b}). Nonetheless, it is written in a way that is more convenient to deal with the conservation of the total EMT.

Applying $\pd{_\lambda}$ on both sides of Eq.~(\ref{16112018d}) and using the property $\superpotential{^a^\lambda^\nu}=-\superpotential{^a^\nu^\lambda}$, we get the identity $\pd{_\lambda}\pd{_\nu}\left(e^{-\varphi/2}e\superpotential{^a^\lambda^\nu} \right)=0$, which leads to the conservation equation 
\begin{equation}
\pd{_\lambda}\left[ e^{(-5\varphi/2)}e\e{^a_\mu}(t^{\lambda\mu}+\theta^{\lambda\mu}-\frac{\Lambda}{\chi}e^{\varphi}g^{\mu\lambda})\right]=0.\label{16112018e}
\end{equation}
Following the standard approach (see, e.g., Ref.~\cite{ANDP:ANDP201200272}), we define the energy-momentum contained within a three-dimensional volume  $V$ as
\begin{equation}
P^a\equiv \int d^3x e^{(-5\varphi/2)}e\e{^a_\mu}(t^{0\mu}+\theta^{0\mu}-\frac{\Lambda}{\chi}e^{\varphi}g^{\mu 0}), \label{16112018f}
\end{equation}
which can be recast with the help of Eq.~(\ref{16112018d}) in the form
\begin{equation}
P^a=-\frac{2}{\chi}\oint dS_j e^{-\varphi/2}e\superpotential{^a^0^j}, \label{16112018g}
\end{equation}
where use of the Gauss' theorem has been made, and $j=1,2,3$ is a coordinate index.

It is straightforward to check that Eqs.~(\ref{16112018b}) and (\ref{16112018d})-(\ref{16112018g}) are covariant under WT. The expression (\ref{16112018g}), in particular, is very similar to the equation (20) in Ref.~\cite{doi:10.1142/S0217732317501139}. However, these equations are not equivalent. The main difference is the fact that the integrand of equation (20) in that reference is not covariant. To compare the two expressions, use the relation $\phi=e^{-\varphi/2}$, where $\phi$ is the same as the one in Eq.~(\ref{25102018a}).

\section{Solutions}\label{17122018b}
 Before we dive into solutions of Eq.~(\ref{16112018d}), an important point here needs to be clarified. If we want, we can recover all of the solutions of GR (TEGR). For this case, we just need to impose the boundary conditions on the metric $\tilde{g}_{\mu\nu}$ (the metric in the teleparallel frame $\tensor{\tilde{e}}{_a^\mu}$, $\tilde{\varphi}=0$) and the solutions would  be those of GR, even if we went to a general Weyl frame after solving the field equations. However, since the theory we are working with is not invariant under the choice of the frame $(\e{_a},\varphi)$ where we impose the boundary conditions, it is more interesting to find solutions with the boundary conditions imposed in a generic Weyl frame $(\e{_a},\varphi\neq 0)$: we call these solutions ``nonequivalent solutions''.  Let us start with some vacuum ones, which can be obtained by demanding that Eq.~(\ref{11072017a}) vanishes. 

\subsection{Plane waves}
Here, it is assumed that in the frame $(\e{_a},\varphi\neq 0)$, the spacetime metric $g$ is given by the metric of the  plane-fronted gravitational waves with parallel rays, known as $pp$-waves, which can be written in the form (see, e.g., Ref.~\cite{griffiths1991colliding})
\begin{equation}
ds^2=2H(u,x,y)du^2+2dudv-dx^2-dy^2, \label{20082017a}
\end{equation}
where $u$ is a null coordinate.

A convenient choice for our tetrad field is
\begin{align}
\theta^{(0)}=\left(H+1\right)\frac{du}{\sqrt{2}}+\frac{dv}{\sqrt{2}},\ \theta^{(1)}=(H-1)\frac{du}{\sqrt{2}}+\frac{dv}{\sqrt{2}},
\nonumber\\
 \theta^{(2)}=dx,\ \theta^{(3)}=dy; \label{20082017b}
\end{align}

\subsubsection{Solution for $\varphi(u,x,y)$} \label{23122018b}
From Eq.~(\ref{20082017b}) and the assumption that $\varphi=\varphi(u,x,y)$, the tensors $\torsion{^a_b_c}$ and $\Sigma_{abc}$ become (the comma indicates partial derivative):
\begin{align}
\torsion{^{(1)}_{(1)}_{(0)}}=\torsion{^{(2)}_{(2)}_{(0)}}=\torsion{^{(3)}_{(3)}_{(0)}}=\torsion{^{(0)}_{(1)}_{(0)}}=
\nonumber\\
=\torsion{^{(2)}_{(1)}_{(2)}}=\torsion{^{(3)}_{(1)}_{(3)}}=\superpotential{_{(1)(1)(0)}}=
\nonumber\\
=\superpotential{_{(2)(2)(0)}}=\superpotential{_{(3)(3)(0)}}=\superpotential{_{(0)(0)(1)}}=
\nonumber\\
=\superpotential{_{(2)(1)(2)}}=\superpotential{_{(3)(1)(3)}}=\frac{\sqrt{2}}{4}\varphi_{,u},
 \label{06122017a}
\end{align}
\begin{align}
\torsion{^{(1)}_{(2)}_{(0)}}=\torsion{^{(0)}_{(1)}_{(2)}}=2\superpotential{_{(1)(0)(2)}}=2\superpotential{_{(0)(1)(2)}}=
\nonumber\\
=\frac{1}{2}H_{,x},
\end{align}
\begin{align}
\torsion{^{(1)}_{(3)}_{(0)}}=\torsion{^{(0)}_{(1)}_{(3)}}=2\superpotential{_{(1)(0)(3)}}=2\superpotential{_{(0)(1)(3)}}=
\nonumber\\
=\frac{1}{2}H_{,y},
\end{align}
\begin{align}
\torsion{^{(3)}_{(3)}_{(2)}}=\superpotential{_{(3)(3)(2)}}=\frac{1}{2}\varphi_{,x},
\end{align}
\begin{align}
\torsion{^{(2)}_{(2)}_{(3)}}=\superpotential{_{(2)(2)(3)}}=\frac{1}{2}\varphi_{,y},
\end{align}
\begin{align}
\torsion{^{(0)}_{(2)}_{(0)}}=\frac{1}{2}\left(H_{,x}-\varphi_{,x} \right),\ \torsion{^{(0)}_{(3)}_{(0)}}=
\nonumber\\
=\frac{1}{2}\left(H_{,y}-\varphi_{,y} \right),
\end{align}
\begin{align}
\torsion{^{(1)}_{(1)}_{(2)}}=\frac{1}{2}\left(H_{,x}+\varphi_{,x} \right),\ \torsion{^{(1)}_{(1)}_{(3)}}=
\nonumber\\
=\frac{1}{2}\left(H_{,y}+\varphi_{,y} \right),
\end{align}
\begin{align}
\superpotential{_{(0)(2)(0)}}=\frac{1}{4}H_{,x}+\frac{1}{2}\varphi_{,x},\ \superpotential{_{(0)(3)(0)}}=
\nonumber\\
=\frac{1}{4}H_{,y}+\frac{1}{2}\varphi_{,y},
\end{align}
\begin{align}
\superpotential{_{(1)(2)(1)}}=\frac{1}{4}H_{,x}-\frac{1}{2}\varphi_{,x},\ \superpotential{_{(1)(3)(1)}}=
\nonumber\\
=\frac{1}{4}H_{,y}-\frac{1}{2}\varphi_{,y}.
\end{align}
These expressions lead to 
\begin{equation}
\calt=\frac{3}{2}\left( \varphi_{,x}^2+\varphi_{,y}^2\right).
\end{equation}

Equating Eqs. (\ref{11072017a}) and (\ref{20072017b}) to zero and denoting them by $\tensor{E}{_c^\lambda}$ and $E_{\varphi}$, respectively, we find that:
\begin{equation}
\tensor{E}{_{(0)}^u}=\sqrt{2}e^{-\varphi}\left(\varphi_{,xx}+\varphi_{,yy}-\frac{1}{4}\varphi_{,x}^2-\frac{1}{4}\varphi_{,y}^2 \right)=0, \label{22082017a}
\end{equation}
which yields
\begin{equation}
\varphi_{,xx}+\varphi_{,yy}=\frac{1}{4}(\varphi_{,x}^2+\varphi_{,y}^2), \label{22082017b}
\end{equation}
and
\begin{equation}
E_{\varphi}=3e^{-\varphi}\left(\frac{1}{2}\varphi_{,x}^2+\frac{1}{2}\varphi_{,y}^2-\varphi_{,xx}-\varphi_{,yy} \right)=0, \label{22082017c}
\end{equation}
which gives
\begin{equation}
\varphi_{,xx}+\varphi_{,yy}=\frac{1}{2}(\varphi_{,x}^2+\varphi_{,y}^2). \label{22082017d}
\end{equation}
Comparing Eq.~(\ref{22082017d}) with (\ref{22082017b}), we see that $\varphi_{,x}^2+\varphi_{,y}^2=0$. As $\varphi$ is real, we must have  $\varphi_{,x},\varphi_{,y}=0$, that is $\varphi=\varphi(u)$. Now, if we calculate the other components of $\tensor{E}{_c^\lambda}$, we find that the only nonzero ones are $\tensor{E}{_{(0)}^v}=-\tensor{E}{_{(1)}^v}$ and the final equation is
\begin{equation}
H_{,xx}+H_{,yy}+ \ddot{\varphi}+\frac{1}{2}\dot{\varphi}^2=0, \label{22082017f}
\end{equation}
where $\dot{\varphi}=d\varphi/du$.

This wave equation will be the same as that of GR for 
\begin{equation}
\ddot{\varphi}+\frac{1}{2}\dot{\varphi}^2=0, \label{22082017g}
\end{equation}
whose solution is
\begin{equation}
\varphi(u)=2\ln(Cu+1)+\bar{C}, \label{22082017h}
\end{equation}
where $C$ and $\bar{C}$ are integration constants.

\subsubsection{Solution for $\varphi(u,v,x,y)$} 
With $\varphi $ also depending on $v$, the components $\torsion{^a_b_c}$ and $\superpotential{_a_b_c}$ differ from the previous case only for Eq.~(\ref{06122017a}), in which case they take the form 
\begin{align}
\torsion{^{(0)}_{(1)}_{(0)}}=\torsion{^{(2)}_{(1)}_{(2)}}=\torsion{^{(3)}_{(1)}_{(3)}}=\superpotential{_{(0)}_{(0)}_{(1)}}=
\nonumber\\
=\superpotential{_{(2)}_{(1)}_{(2)}}=\superpotential{_{(3)}_{(1)}_{(3)}}=\frac{\sqrt{2}}{4}\left[\varphi_{,u}-(H+1)\varphi_{,v}\right], \label{23122017a}
\end{align}
\begin{align}
\torsion{^{(1)}_{(1)}_{(0)}}=\torsion{^{(2)}_{(2)}_{(0)}}=\torsion{^{(3)}_{(3)}_{(0)}}=\superpotential{_{(1)}_{(1)}_{(0)}}=
\nonumber\\
=\superpotential{_{(2)}_{(2)}_{(0)}}=\superpotential{_{(3)}_{(3)}_{(0)}}=\frac{\sqrt{2}}{4}\left[\varphi_{,u}-(H-1)\varphi_{,v}\right]. \label{23122017b}
\end{align}
In this case, we have $\torsion{}=(3/2)(\varphi_{,x}^2+\varphi_{,y}^2)+3\varphi_{,v}^2H-3\varphi_{,u}\varphi_{,v}$.

Let us start from the simplest equations, which corresponds to $\tensor{E}{_{(2)}^y}=0$, $\tensor{E}{_{(2)}^u}=0$ and $\tensor{E}{_{(3)}^u}=0$:
\begin{equation}
\varphi_{,y}\varphi_{,x}+2\varphi_{,xy}=0, \label{31102017a}
\end{equation}
\begin{equation}
\varphi_{,x}\varphi_{,v}+2\varphi_{,vx}=0, \label{31102017b}
\end{equation}
\begin{equation}
\varphi_{,y}\varphi_{,v}+2\varphi_{,vy}=0. \label{31102017c}
\end{equation}
Integration of Eqs. (\ref{31102017b}) and (\ref{31102017c}) gives
\begin{equation}
2\left( e^{\varphi/2}\right)_{,v}=C_1(u,v),
\end{equation}
while Eq.~(\ref{31102017a}) can be integrated with respect to $x$ and also $y$ to yield
\begin{equation}
2\left( e^{\varphi/2}\right)_{,y}=C_2(u,v,y),
\end{equation}
\begin{equation}
2\left( e^{\varphi/2}\right)_{,x}=C_3(u,v,x);
\end{equation}
$C_1(u,v)$, $C_2(u,v,y)$ and $C_3(u,v,x)$ are functions of integration. It follows immediately from the last three equations that we must separate the dependence of $x$, $y$ and $v$; that is, $e^{\varphi/2}=f_1(u,v)+f_2(u,x)+f_3(u,y)$. Therefore, the solution to them is
\begin{equation}
\varphi(u,v,x,y)=2\ln\left[f_1(u,v)+f_2(u,x)+f_3(u,y) \right]. \label{8092017a}
\end{equation}
From Eq.  $\tensor{E}{_{(2)}^v}=0$ we get
\begin{equation}
-f_2(u,x)_{,ux}+H(u,x,y)_{,x}f_1(u,v)_{,v}=0.
\end{equation}
By assuming that $H(u,x,y)_{,x}\neq 0$, we obtain
\begin{equation}
f_1(u,v)=f(u)v+C, \label{8092017b}
\end{equation}
where $C$ is a constant. Now, if we use Eqs. (\ref{8092017b}) and (\ref{8092017a}) to calculate the equations $\tensor{E}{_{(2)}^x}=0$ and $\tensor{E}{_{(3)}^y}=0$ , we will obtain polynomial equations for $v$. From the coefficients  of terms with the same power of $v$, we get
\begin{equation}
f_2(u,x)_{,xx}=f_3(u,y)_{,yy}=-\dot{f}(u), \label{31102017d}
\end{equation}
which yields
\begin{equation}
f_2(u,x)=F(u)+D(u)x-\frac{1}{2}\dot{f}(u)x^2, \label{31102017e}
\end{equation}
\begin{equation}
f_3(u,y)=G(u)+B(u)y-\frac{1}{2}\dot{f}(u)y^2. \label{31102017f}
\end{equation}
Substituting (\ref{8092017b})-(\ref{31102017f}) into (\ref{8092017a}), we obtain
\begin{eqnarray}
\varphi(u,v,x,y)&=&2\ln\biggl[E(u)+f(u)v+D(u)x+B(u)y
\nonumber\\
&&-\frac{1}{2}\dot{f}(u)\left(x^2+y^2 \right) \biggr], \label{8092017c}
\end{eqnarray}
where $E(u)=C+F(u)+G(u)$.  Using this expression in equations $\tensor{E}{_{(2)}^v}=0$ and $\tensor{E}{_{(3)}^v}=0$ gives
\begin{equation}
-\dot{D}(u)+\ddot{f}(u)x+H(u,x,y)_{,x}f(u)=0,
\end{equation}
\begin{equation}
-\dot{B}(u)+\ddot{f}(u)y+H(u,x,y)_{,y}f(u)=0.
\end{equation}
Integrating with respect to $x$ and $y$, respectively, we obtain
\begin{equation}
-\dot{D}(u)x+\ddot{f}(u)\frac{x^2}{2}+H(u,x,y)f(u)=h_1(u,y), \label{08092017d}
\end{equation}
\begin{equation}
-\dot{B}(u)y+\ddot{f}(u)\frac{y^2}{2}+H(u,x,y)f(u)=h_2(u,x),
\end{equation}
which can be manipulated to give a function of $u$ only:
\begin{align}
-\dot{D}(u)x+\ddot{f}(u)\frac{x^2}{2}+h_2(u,x)=&-\dot{B}(u)y+\ddot{f}(u)\frac{y^2}{2}
\nonumber\\
&+h_1(u,y) \equiv k(u).
\end{align}
Note that the first equality holds only if the dependences on $x$ and $y$ disappear, which justifies the definition of the function $k(u)$. From the second equality, we can isolate $h_1(u,y)$ to write it in terms of $k(u)$ and then substitute $h_1$ into (\ref{08092017d}) to obtain
\begin{equation}
H(u,x,y)=\frac{k(u)}{f(u)}+\frac{\dot{D}(u)}{f(u)}x+\frac{\dot{B}(u)}{f(u)}y-\frac{\ddot{f}(u)}{2f(u)}(x^2+y^2), \label{8092017e}
\end{equation}
where  $f(u)\neq 0$ has been assumed. The function $k(u)$ cannot be arbitrary. By using (\ref{8092017c}) and (\ref{8092017e}) in $\tensor{E}{_{(0)}^u}=0$, we find that $k$ has to satisfy the relation
\begin{equation}
\frac{k(u)}{f(u)}=-\frac{\dot{f}(u)}{f(u)^2}E(u)+\frac{\dot{E}(u)}{f(u)}-\frac{D(u)^2+B(u)^2}{2f(u)^2}. \label{02112017a}
\end{equation}
It is straightforward to verify that all field equations are satisfied after imposing this condition.

Finally, we can conclude that the solutions to the field equations for the case $\phi(u,v,x,y)$ are (\ref{8092017c}) and (\ref{8092017e}) with the restriction (\ref{02112017a}); functions $E(u)$, $D(u)$, $B(u)$, and $f(u)$ (except for $f\neq 0$) remain arbitrary.

\subsection{Spherically symmetric solutions}
The nonequivalent solution for the Spherically symmetric case can be obtained in the following way. Let us assume that the spacetime metric obeys this symmetry and at the same time we have a nonvanishing Weyl field. In this case, we assume that
\begin{equation}
\theta^a=(e^{\nu(r)/2}dt,e^{\lambda(r)/2}dr,rd\theta,r\sin\theta d\phi) \label{02112018b}
\end{equation}
and search for a solution of Eq.~(\ref{11072017a}) with a nontrivial $\varphi(r)$.

%(\ref{02112018a})
From Eqs.~(\ref{29102017a}), (\ref{20072017a})-(\ref{09112017d}),  we get
\begin{align}
\torsion{^{(2)}_{(1)(2)}}=\torsion{^{(3)}_{(1)(3)}}=\frac{1}{2}\frac{e^{-\lambda/2}}{r}(2-\varphi' r),
\nonumber\\
\torsion{^{(0)}_{(0)(1)}}=\frac{1}{2}e^{-\lambda/2}(\varphi'-\nu'),
\nonumber\\
\torsion{^{(3)}_{(2)(3)}}=\frac{1}{r}\cot\theta, \label{20112018a}
\end{align}
\begin{align}
\superpotential{_{(2)(2)(1)}}=\superpotential{_{(3)(3)(1)}}=\frac{1}{4}\frac{e^{-\lambda/2}}{r}(2r\varphi'-r\nu'-2),
\nonumber\\
\superpotential{_{(0)(0)(2)}}=\superpotential{_{(1)(2)(1)}}=\frac{1}{2}\torsion{^{(3)}_{(2)(3)}},
\nonumber\\
\superpotential{_{(0)(0)(1)}}=\torsion{^{(2)}_{(1)(2)}}, \label{20112018b}
\end{align}
\begin{align}
\torsion{}=\frac{e^{-\lambda}}{2r^2}[3r^2(\varphi')^2-2r^2\varphi'\nu'-8r\varphi'+4r\nu'+4], 
\label{21122018a}\\
e=e^{(\nu+\lambda)/2}r^2\sin\theta,
\label{20112018c}
\end{align}
where the prime denotes derivation with respect to $r$. Now we can use these expressions to calculate Eq.~(\ref{11072017a}):
\begin{align}
\tensor{E}{_{(0)}^t}=&\frac{1}{2}e^{-(\varphi+\lambda/2)}\sin\theta \biggl[ 4r^2\varphi''-r^2(\varphi')^2+8r\varphi'
\nonumber\\
&-2r^2\lambda'\varphi'+4r\lambda'+4e^{\lambda}-4 \biggr], \label{01102017a}
\end{align}
\begin{align}
\tensor{E}{_{(1)}^r}=&-\frac{1}{2}e^{-\varphi +\nu/2-\lambda}\sin\theta[3r^2(\varphi')^2-8r\varphi'
\nonumber\\
&-2r^2\nu'\varphi'+4r\nu'-4e^{\lambda}+4 ], \label{01102017b}
\end{align}
\begin{align}
\tensor{E}{_{(2)}^{\theta}}=&\frac{1}{2}e^{-\varphi +(\nu-\lambda)/2}\sin\theta\biggl[ 4r\varphi''-r(\varphi')^2+4\varphi'
\nonumber\\
&+2r(\nu'-\lambda')\varphi'-2r\nu''-r(\nu')^2-2\nu'
\nonumber\\
&+r\lambda'\nu'+2\lambda' \biggr], \label{01102017c}
\end{align}
%\begin{align} % Retirei, pois não é citada em lugar nenhum e não é necessária
%E_{\phi}=&\frac{1}{2}e^{-\varphi +(\nu-\lambda)/2}\sin\theta\biggl[-6r^2\varphi''+3r^2(\varphi')^2
%\nonumber\\
%&-12r\varphi'-3r^2(\nu'-\lambda')\varphi'+2r^2\nu''+r^2(\nu')^2
%\nonumber\\
%& +4r\nu'-r^2\lambda'\nu'-4r\lambda'-4e^{\lambda}+4\biggr], \label{01102017f}
%\end{align}
and $\tensor{E}{_{(2)}^{\theta}}=\sin\theta\tensor{E}{_{(3)}^{\phi}}$. Adding $\tensor{E}{_{(0)}^t}$ to $-e^{(\lambda-\nu)/2}\tensor{E}{_{(1)}^r}$ and simplifying, we get
\begin{equation}
(\lambda+\nu)'(\frac{1}{r}-\frac{1}{2}\varphi') +\varphi''+\frac{1}{2}(\varphi')^2=0. \label{01102017g}
\end{equation}

\subsubsection{Nonequivalent solution} \label{23122018c} %2/1/2019
For simplicity, let us assume that $\varphi''+\frac{1}{2}(\varphi')^2=0$. This equation is similar to Eq.~(\ref{22082017g}) and its solution can be written as 
\begin{equation}
\varphi=2\ln(Cr+1)+\bar{C}. \label{02112018c}
\end{equation}
 With this assumption, Eq.~(\ref{01102017g}) yields  $(\lambda+\nu)'=0$ (we choose $\lambda=-\nu$). Substituting this result and (\ref{02112018c}) in Eq.~(\ref{01102017c}), we obtain
\begin{equation}
(1+Cr)^2\left[\nu''+(\nu')^2\right]+\frac{2(1-C^2r^2)}{r}\nu'+2C(C-\frac{2}{r})=0,\label{02112018d}
\end{equation}
whose solution is
\begin{align}
\nu(r)=\ln\biggl(\frac{C_1}{r}+2CC_1-C_2+C(CC_1-2C_2)r
\nonumber\\
-C^2C_2r^2 \biggr), \label{02112018f}
\end{align}
where $C_1$ and $C_2$ are integration constants. Substituting this expression into Eq.~(\ref{01102017a}), we obtain the constraint $C_2=-1-CC_1$, which allows us to write
\begin{align}
\nu(r)=&\ln\biggl(1+\frac{C_1}{r}+3CC_1+C(3CC_1+2)r
+C^2(1+CC_1)r^2 \biggr). \label{02012018a}
\end{align}
For $C=0$ ($\varphi=$constant), we get $\nu(r)=\ln(1+C_1/r)$, which is the Schwarzschild case for $C_1<0$, as expected. On the other hand, for  $C\neq 0$ (nontrivial $\varphi$), we have a family of solutions that includes other kinds of solutions. As an example,  we have the case $CC_1=-2/3$, which implies $\nu=\ln\left( -1-2/(3Cr)+C^2r^2/3 \right)$. Nonetheless, this does not prove that there is no solution that embraces the Schwarzschild one with a nontrivial $\varphi$ as a particular case, since we have not considered the general solution in Eq.~(\ref{01102017g}). We will see such a solution in the next section.

It is interesting to note that the case $C_1=0$, which yields $\nu=2\ln(1+Cr)$, is equivalent to the Minkowski spacetime. To show this, we need to go to the teleparallel frame. For simplicity, let us take $\bar{C}=0$, in which case we have $\varphi=2\ln(1+Cr)=\nu$.  From  Eq.~(\ref{08112018a}), we find that $\tilde{g}_{\mu\nu}=diag(1,-e^{-2\nu},-r^2e^{-\nu},-r^2e^{-\nu}\sin^2\theta)$. If we change the coordinate $r$ to $R=r/(1+Cr)$, we will find that this metric is the Minkowski metric written in the spherical coordinate system $(t,R,\theta,\phi)$.

\subsubsection{The teleparallel equivalent solution}\label{23122018a}

Let us seek an expression that ensures that the solution of the field equations in the teleparallel frame ($\tensor{\tilde{e}}{_a^\mu}$ and $\tilde{\varphi}=0$) is the  Schwarzschild one. In doing so, consider the metric tensor in each frame:
\begin{equation}
ds^2=e^{\nu(r)}dt^2-e^{\lambda(r)}dr^2-r^2d\Omega^2, \label{08112018b}
\end{equation}
\begin{equation}
d\tilde{s}^2=e^{\tilde{\nu}(\tilde{r})}dt^2-e^{\tilde{\lambda}(\tilde{r})}d\tilde{r}^2-\tilde{r}^2d\Omega^2, \label{08112018c}
\end{equation}
where 
\begin{align}
\tilde{\nu}(\tilde{r})=\ln(1-2m/\tilde{r}),\ \tilde{\lambda}(\tilde{r})=-\ln(1-2m/\tilde{r}). \label{19112018b}
\end{align}
Note that $d\Omega^2$ was chosen to be the same for both frames because we are going to use $\varphi$ depending only on $r$. Furthermore, these metrics are related to each other through $d\tilde{s}^2=e^{2\theta}ds^2=e^{-\varphi}ds^2$ (when the coordinate systems are different, we must use the invariant form, rather than $\tilde{g}_{\mu\nu}=e^{-\varphi}g_{\mu\nu}$). This relation yields 
\begin{equation}
\tilde{r}=re^{-\varphi/2}, \label{19112018a}
\end{equation} 
which leads to $d\tilde{r}=(1-r\varphi'/2)e^{-\varphi/2}dr$. This coordinate change is clearly problematic for $\varphi=2\ln r$ and, for the sake of simplicity, we will deal only with the cases $1-r\varphi'/2>0$.   Substituting the relation between  $d\tilde{r}$ and $dr$ into Eq.~(\ref{08112018c}) and using $d\tilde{s}^2=e^{-\varphi}ds^2$ again, one arrives at
\begin{equation}
\tilde{\nu}(\tilde{r})=\nu(r)-\varphi(r),\ \tilde{\lambda}(\tilde{r})=\lambda(r)-2\ln(1-\frac{r}{2}\varphi'). \label{19112018e}
\end{equation}
Using the fact that $\tilde{\nu}+\tilde{\lambda}=0$, we find  $\lambda=-\nu+\varphi+2\ln(1-\frac{r}{2}\varphi')$ [one can easily verify that Eq.~(\ref{01102017g}) is satisfied by this expression]. Making use of this result in Eq.~(\ref{01102017b}), we obtain the equation
\begin{align}
-3r\varphi'+r\varphi'e^{-\nu+\varphi}+2r\nu'+2-2e^{-\nu+\varphi}=0,
\end{align}
whose solution is
\begin{align}
\nu=\frac{3}{2}\varphi+\ln\left(\frac{C_1}{r}+e^{-\varphi/2}\right). \label{19112018c}
\end{align}
Therefore,
\begin{align}
\lambda=-\frac{1}{2}\varphi+\ln\left[\frac{(1-\frac{r}{2}\varphi')^2}{\frac{C_1}{r}+e^{-\varphi/2}} \right]. \label{19112018d}
\end{align}
Substitution of Eqs.~(\ref{19112018c})-(\ref{19112018d}) into Eqs.~(\ref{01102017a})-(\ref{01102017c}) shows that they solve the field equations regardless of Eq.~(\ref{19112018b}). To finally connect this solution to the Schwarzschild one, we need to substitute Eq.~(\ref{19112018c}) or Eq.~(\ref{19112018d}) into Eq.~(\ref{19112018e}) and use Eq.~(\ref{19112018b}). This procedure yields $C_1=-2m$. Notice that $\varphi$ remains arbitrary, except for the restriction $1-r\varphi'/2>0$. This solution belongs to a subset of solutions that are equivalent to the Schwarzschild one (in the formalism considered here) and are spherically symmetric, but not necessarily asymptotically flat in the frame ($\e{_a},\varphi(r)\neq 0$). This is the solution we would have obtained if we had solved the field equations directly  in the teleparallel frame with the assumptions that the spacetime is asymptotically flat      and spherically symmetric, and then applied a WT with $\theta$ given by $\theta=-\varphi(r)/2$.

To exemplify the equivalence of this solution with the Schwarzschild one, let us calculate Eq.~(\ref{16112018g})  for the surface $t$ and $r$ constant. For this we use
\begin{equation}
\e{^c_\lambda}=\left( 
\begin{array}{cccc}
e^{\frac{\nu}{2}}&0&0&0\\
0&e^{\frac{\lambda}{2}}\sin\theta\cos\phi & r\cos\theta\cos\phi & -r\sin\theta\sin\phi\\
0&e^{\frac{\lambda}{2}}\sin\theta\sin\phi & r\cos\theta\sin\phi & r\sin\theta\cos\phi\\
0&e^{\frac{\lambda}{2}}\cos\theta & -r\sin\theta & 0
\end{array}
\right), \label{21112018a}
\end{equation}
which can be inverted to
\begin{equation}
\e{_c^\lambda}=\left( 
\begin{array}{cccc}
e^{-\frac{\nu}{2}}&0&0&0\\
0&e^{-\frac{\lambda}{2}}\sin\theta\cos\phi & \frac{\cos\theta\cos\phi}{r} & -\frac{\sin\phi}{r\sin\theta}\\
0&e^{-\frac{\lambda}{2}}\sin\theta\sin\phi & \frac{\cos\theta\sin\phi}{r}  & \frac{\cos\phi}{r\sin\theta} \\
0&e^{-\frac{\lambda}{2}}\cos\theta & -\frac{\sin\theta}{r} & 0
\end{array}
\right). \label{22112018a}
\end{equation}
The reason why we must not use the tetrad field (\ref{02112018b}) to calculate a quantity that depends on the tetrad field, such as $P^a$, was given in the paragraph right after Eq.~(\ref{05112017a}).

As always, we can read off the components of the torsion tensor from Eqs.~(\ref{29102017a}) and (\ref{05112017a}):
\begin{align}
\torsion{^{(0)}_{(0)(1)}}=\frac{e^{-\lambda/2}}{2}\sin\theta\cos\phi(\varphi'-\nu'),
\nonumber\\
 \torsion{^{(0)}_{(0)(2)}}=\frac{e^{-\lambda/2}}{2}\sin\theta\sin\phi(\varphi'-\nu'),
\nonumber\\ 
   \torsion{^{(0)}_{(0)(3)}}=\frac{e^{-\lambda/2}}{2}\cos\theta(\varphi'-\nu'),
\nonumber\\
\torsion{^{(1)}_{(1)(2)}}=\frac{e^{-\lambda/2}}{2r}\sin\theta\sin\phi(r\varphi'-2+2e^{\lambda/2}),
\nonumber\\
\torsion{^{(1)}_{(1)(3)}}=\frac{e^{-\lambda/2}}{2r}\cos\theta(r\varphi'-2+2e^{\lambda/2}),
\nonumber\\
\torsion{^{(2)}_{(1)(2)}}=-\frac{e^{-\lambda/2}}{2r}\sin\theta\cos\phi(r\varphi'-2+2e^{\lambda/2}),
\nonumber\\
\torsion{^{(2)}_{(2)(3)}}=\torsion{^{(1)}_{(1)(3)}},\ \torsion{^{(3)}_{(1)(3)}}=\torsion{^{(2)}_{(1)(2)}},
\nonumber\\
\torsion{^{(3)}_{(2)(3)}}=-\torsion{^{(1)}_{(1)(2)}}.
\end{align}
Since we want $\superpotential{^a^0^1}=e^{-\nu/2}\e{_c^1}\superpotential{^a^{(0)}^c}$, we need only the following components:
\begin{align}
\superpotential{^{(0)(0)(1)}}=\torsion{^{(2)}_{(2)(1)}},\, \superpotential{^{(0)(0)(2)}}=\torsion{^{(1)}_{(1)(2)}},
\nonumber\\
\superpotential{^{(0)(0)(3)}}=\torsion{^{(1)}_{(1)(3)}}
\end{align}
and $\superpotential{^{(j)(0)}^c}=0$. Thus, we have $\superpotential{^a^0^1}=0$ for $a\neq (0)$ and $\superpotential{^{(0)}^0^1}=e^{-(\nu/2+\lambda)}(r\varphi'-2+2e^{\lambda/2})/(2r)$. Using this expression and the determinant $e$, given by Eq.~(\ref{20112018c}), into Eq.~(\ref{16112018g}), we obtain
\begin{align}
P^{(0)}=\frac{4\pi}{-\chi}re^{-(\varphi+\lambda)/2}(r\varphi'-2+2e^{\lambda/2}) \label{23112018a}
\end{align}
and $P^{(j)}=0$. Finally, substituting Eq.~(\ref{19112018d}) into this expression and using Eq.~(\ref{19112018a}), we arrive at $P^{(0)}=(c^4/G)\tilde{r}[1-(1-2m/\tilde{r})^{1/2}]$  (remember that $\chi=-8\pi G/c^4$ and $C_1=-2m$), which is exactly the Schwarzschild case in the TEGR, as expected.

\section{Cosmology}\label{17122018c}
The Lagrangian density of a perfect fluid in GR can be written as $\call^{\textrm{\tiny GR}}_M=2e\rho\left(|J|/e,s\right)$ \cite{0264-9381-10-8-017}, where $\rho$ is the energy density, $s$ is the entropy per particle and $|J|$ is the magnitude of the contravariant vector density defined as $J^{\mu}=e n u^{\mu}$ ($n$ is the particle number density and $u^{\mu}$ is the fluid $4$-velocity). Since the variable $s$ is seen as independent of $\e{^c_\lambda}$, we will omit it.

Exchanging $\tensor{e}{^c_\lambda}$ for $e^{-\varphi/2}\tensor{e}{^c_\lambda}$ gives
\begin{equation}
\call_M=2ee^{-2\varphi}\rho\left(e^{(3/2)\varphi}\frac{|J|}{e}\right), \label{12042018a}
\end{equation}
where we have exchanged $|J|$ for $e^{-\varphi/2}|J|$, since $|J|=\sqrt{J^{\mu}J^{\nu}g_{\mu\nu}}$ and $J^{\mu}$ is considered  to be independent of $\e{^c_\mu}$. Keep in mind that in Ref.~\cite{0264-9381-10-8-017} the author consider $J^{\mu}$ and $g_{\mu\nu}$ as independent variables, while here we are taking $J^{\mu}$ and $\e{^c_\mu}$ instead. In short, we have $\call_M=\call_M(\e{^c_\lambda},\varphi,J^{\lambda})$.

From Eqs.~(\ref{11112018a}) and (\ref{12042018a}), we obtain
\begin{equation}
\theta^{\mu\nu}=e^{\varphi}\left[ (\rho+p)u^{\mu}u^{\nu}-pg^{\mu\nu}\right], \label{11112018c}
\end{equation}
where $p\equiv n\frac{\partial \rho}{\partial n}-\rho$, as in Ref.~\cite{0264-9381-10-8-017}. Notice that, since $\theta^{\mu\nu}$ is invariant under WT (keep in mind that $u^{\mu}=\e{_{(0)}^\mu}$), we could have obtained Eq.~(\ref{11112018c}) by  following the coupling prescription directly in the EMT of a perfect fluid in GR. Furthermore, both $\rho$ and $p$ are also invariant.

Substitution of Eqs.~(\ref{25082018h}) and (\ref{11112018c}) into (\ref{11112018b}) gives
\begin{equation}
\frac{\delta \call_g}{\delta \tensor{e}{^c_\lambda}}-2\Lambda ee^{-2\varphi}\tensor{e}{_c^\lambda}+2\chi ee^{-2\varphi}\left[ (\rho+p)u_c u^{\lambda}-p\tensor{e}{_c^\lambda} \right]=0. \label{12042018b}
\end{equation}
Recall that the first term in this equation is given by Eq.~(\ref{11072017a}). Alternatively, one can use this field equation in the form of Eq.~(\ref{16112018d}).

\subsection{The energy of the universe} \label{11122018c}
To calculate the energy of the universe we assume that the spacetime is given by the Robertson-Walker line element:
\begin{equation}
ds^2=dt^2-a(t)^2\frac{\left[d\rb^2+\rb^2(d\theta^2+\sin^2\theta d\phi^2) \right]}{[1+\frac{1}{4}k \rb^2]^2}, \label{27092018c}
\end{equation}
where $k=0,\pm 1$. For the time being, we do not specify whether this condition holds on the teleparallel frame (whether we want an equivalent solution).

To perform the calculation, we use
%\begin{widetext}
\begin{equation}
\tensor{e}{^a_\mu}=\left(
\begin{array}{cccc}
1 & 0 & 0 & 0
\\
0 & f\sin\theta\cos\phi & f\rb\cos\theta\cos\phi & -f\rb\sin\theta\sin\phi
\\
0 & f\sin\theta\sin\phi & f\rb\cos\theta\sin\phi & f\rb\sin\theta\cos\phi
\\
0 & f\cos\theta & -f\rb\sin\theta & 0
\end{array}\right),  \label{27092018d}
\end{equation}
%\end{widetext}
where $f(t,\rb)=a(t)/(1+k\rb^2/4)$. The main advantage of using this tetrad is that it becomes the Cartesian basis of $1$-form, that is $\theta^a=\tensor{e}{^a_\mu}dx^{\mu}=(dt,dx,dy,dz)$, when $a$ is constant and $k=0$. Therefore, the torsion tensor will also vanish in Minkowski spacetime. This feature is irrelevant for the field equations, which are covariant under LLT, but it is fundamental for the calculation of the energy of the Universe.

For simplicity, let us assume that $\varphi=\varphi(t)$. Using this assumption and the tetrad (\ref{27092018d}) in Eqs.~(\ref{29102017a}), (\ref{20072017a}), and (\ref{09112017d}) [remember that we are using the values in Eq. (\ref{02112018a})], we obtain:
\begin{align}
\torsion{^{(1)}_{(0)(1)}}=\torsion{^{(2)}_{(0)(2)}}=\torsion{^{(3)}_{(0)(3)}}=\frac{\dot{a}}{a}-\frac{1}{2}\dot{\varphi},
\nonumber\\
\torsion{^{(1)}_{(1)(2)}}=-\torsion{^{(3)}_{(2)(3)}}=\frac{\rb k}{2a}\sin\theta\sin\phi, 
\nonumber\\
\torsion{^{(2)}_{(1)(2)}}=\torsion{^{(3)}_{(1)(3)}}=-\frac{\rb k}{2a}\sin\theta\cos\phi,
\nonumber\\
\torsion{^{(1)}_{(1)(3)}}=\torsion{^{(2)}_{(2)(3)}}=\frac{\rb k}{2a}\cos\theta, \label{09122018a}
\end{align}
\begin{align}
\superpotential{^{(0)(0)(1)}}=2\superpotential{^{(2)(1)(2)}}=2\superpotential{^{(3)(1)(3)}}=-\torsion{^{(2)}_{(1)(2)}},
\nonumber\\
\superpotential{^{(1)(0)(1)}}=\superpotential{^{(2)(0)(2)}}=\superpotential{^{(3)(0)(3)}}=\torsion{^{(1)}_{(0)(1)}},
\nonumber\\
\superpotential{^{(0)(0)(2)}}=-2\superpotential{^{(1)(1)(2)}}=2\superpotential{^{(3)(2)(3)}}=\torsion{^{(1)}_{(1)(2)}},
\nonumber\\
\superpotential{^{(0)(0)(3)}}=-2\superpotential{^{(1)(1)(3)}}=-2\superpotential{^{(2)(2)(3)}}=\torsion{^{(1)}_{(1)(3)}}, \label{09122018b}
\end{align}
\begin{align}
\calt=-6\frac{\dot{a}^2}{a^2}+\frac{\rb^2k^2}{2a^2}+6\dot{\varphi}\frac{\dot{a}}{a}-\frac{3}{2}\dot{\varphi}^2, \label{09122018c}
\end{align}
and also 
\begin{equation}
e=a^3\rb^2\sin\theta/(1+k\rb^2/4)^3. \label{09122018d}
\end{equation}
From Eq.~(\ref{16112018g}), we see that to find the energy-momentum within a three-dimensional sphere with $\rb$ constant we need only the components $\superpotential{^a^0^1}$. From Eqs.~(\ref{27092018c}), (\ref{27092018d}), and (\ref{09122018b}), we get:
\begin{align}
\superpotential{^{(0)}^0^1}=\frac{k\rb}{2a^2}(1+k\rb^2/4),\ \superpotential{^{(1)}^0^1}=h(t,\rb)\sin\theta\cos\phi,
\nonumber\\
\superpotential{^{(2)}^0^1}=h(t,\rb)\sin\theta\sin\phi,\, \superpotential{^{(3)}^0^1}=h(t,\rb)\cos\theta,
\nonumber\\
h(t,\rb)=\frac{(1+k\rb^2/4)}{a}(\dot{a}/a-\dot{\varphi}/2). \label{09122018e}
\end{align}
Finally, we use Eqs.~(\ref{09122018d}) and (\ref{09122018e}) in Eq.~(\ref{16112018g}) to obtain 
\begin{equation}
E=\frac{c^4}{2G}\frac{\rb^3}{(1+k\rb^2/4)^2}ka(t)e^{-\varphi(t)/2}, \label{09122018f}
\end{equation}
and $P^{(j)}=0$, where $\chi=-8\pi G/c^4$ has been used.

This result is different from the one obtained in Ref.~\cite{doi:10.1142/S0217732317501139} not only because
the authors use a slightly different definition for $P^a$, which is responsible for the factor $\phi^2$ there (here we have $\phi=e^{-\varphi/2}$), but also because the tetrad used here is different. Furthermore, the total energy of the universe given by Eq.~(\ref{09122018f}) vanishes  for $k=0$, while the one in Ref.~\cite{doi:10.1142/S0217732317501139} does not (this happens because their tetrad field is not holonomic in the Minkowski spacetime).

One may write Eq.~(\ref{09122018f}) in the same coordinate system that is used in Ref.~\cite{doi:10.1142/S0217732317501139} by using the relation $\rb=2(1-\sqrt{1-kr^2})/(kr)$. In this new coordinate, this equation takes on the form
\begin{equation}
E=\frac{c^4}{G}(1-\sqrt{1-kr^2})r a(t)e^{-\varphi(t)/2}. \label{09122018g}
\end{equation}
For $\varphi=0$, we have the total energy of the universe as predicted by the TEGR.

\subsection{The gravitational energy of the universe} \label{11122018d}

Based on the definition (\ref{16112018f}), it is natural to assume that the gravitational energy within the spherical volume $V$ is given by\footnote{Note that, in this approach, the cosmological constant is being treated as a matter field, i.e., it is out of the gravitational energy. }
\begin{equation}
E_g=\int_V dx^3 e^{-5\varphi/2}e t^{0(0)}. \label{11122018a}
\end{equation}
%27092018c 27092018d 09122018a 09122018b 09122018c
In turn, from Eq.~(\ref{16112018b}) and (\ref{27092018c})-(\ref{09122018c}), we have
\begin{align}
t^{0(0)}=\frac{e^{2\varphi}}{2\chi}\left(6\frac{\dot{a}^2}{a^2}+\frac{k^2\rb^2}{2a^2}-6\dot{\varphi}\frac{\dot{a}}{a}+\frac{3}{2}\dot{\varphi}^2\right). \label{11122018b}
\end{align}
Denoting the integrand of Eq.~(\ref{11122018a}) by $\varrho_g$ and using Eqs.~(\ref{09122018d}) and (\ref{11122018b}), we obtain
\begin{equation}
\varrho_g=\frac{\left[F_1(t)+F_2(t)\rb^2\right]\rb^2\sin\theta}{(1+k\rb^2/4)^3},  \label{11122018e}
\end{equation}
where
\begin{align}
F_1(t)=\frac{a^3e^{-\varphi/2}}{2\chi}\left(6\frac{\dot{a}^2}{a^2}-6\dot{\varphi}\frac{\dot{a}}{a}+\frac{3}{2}\dot{\varphi}^2\right),  \label{11122018f}
\\
F_2(t)=\frac{a^3e^{-\varphi/2}}{2\chi}\frac{k^2}{2a^2}.  \label{11122018g}
\end{align}
Thus the gravitational energy is
\begin{equation}
E_g=4\pi\left[F_1(t)\int_0^{\rb}\frac{u^2du}{(1+\frac{ku^2}{4})^3}+F_2(t)\int_0^{\rb}\frac{u^4du}{(1+\frac{ku^2}{4})^3} \right]. \label{12122018a}
\end{equation}

\subsubsection{The case $k=0$}
For $k=0$, we have $F_2=0$ and $\rb=r$. Therefore, Eq.~(\ref{12122018a}) becomes
\begin{equation}
E_g=\frac{c^2}{2G}a^3r^3\left( -\frac{\dot{a}^2}{a^2}+\dot{\varphi}\frac{\dot{a}}{a}-\frac{1}{4}\dot{\varphi}^2\right) e^{-\varphi/2}, \label{12122018d}
\end{equation}
where, again, we have recovered $c$. Notice that, since the total energy inside $V$ vanishes for $k=0$, the energy of the matter fields plus the one associated with the cosmological constant must be equal to $-E_g$ [see, e.g., Eq.~(\ref{16112018f})]. Note also that $\varphi$ has been treated as a geometric field.
 
For $\varphi=0$, we have
\begin{equation}
E_g=-\frac{c^2}{2G}H^2a^3r^3, \label{12122018b}
\end{equation}
where $H\equiv \dot{a}/a$. This is the gravitational energy predicted by  the TEGR when $k=0$.

\subsubsection{The case $k=1$}
For this case, Eq.~(\ref{12122018a}) can be written in the form
\begin{equation}
E_g=4\pi\left[F_1(t)I_1(\rb)+F_2(t)I_2(\rb) \right] \label{12122018e}
\end{equation}
with $F_1$ and $F_2$ given by Eqs.~(\ref{11122018f})-(\ref{11122018g}) (also $k=1$), and 
\begin{align}
I_1(\rb)=-\frac{\rb h_-}{2h_+^2}+\arctan(\rb/2), \label{12122018f}
\\
I_2(\rb)=-\frac{\rb(12+5\rb^2)}{2h_+^2}+12\arctan(\rb/2), \label{12122018g}
\end{align}
where $h_{\pm}\equiv (1\pm\rb^2/4)$.

\subsubsection{The case $k=-1$}
The case $k=-1$ allows us to put Eq.~(\ref{12122018a}) in the form
\begin{equation}
E_g=4\pi\left[F_1(t)I_3(\rb)+F_2(t)I_4(\rb) \right], \label{12122018h}
\end{equation}
where
\begin{align}
I_3=\frac{\rb h_+}{2h_-^2}+\frac{1}{2}\ln\left|\frac{1-\rb/2}{1+\rb/2}\right|, \label{12122018i}
\\
I_4=\frac{10\rb h_+-16\rb}{h_-^2}-6\ln\left|\frac{1-\rb/2}{1+\rb/2}\right|. \label{12122018j}
\end{align}
As in the case of vanishing curvature, we can find the TEGR versions of the gravitational energy within the three-dimensional volume V by taking $\varphi=0$.

\subsection{The $\Lambda$ energy } \label{12122018l}
One may or may not be interested in assuming a nonvanishing $\Lambda$ even in the presence of $\varphi$. After all, based on the approach considered here, the scalar field $\varphi$ does not have to be related to a cosmological constant: the dark energy may just come from both.

From Eq.~(\ref{16112018f}), we see that the energy associated with $\Lambda$ is
\begin{equation}
E_{\Lambda}=-\frac{\Lambda}{\chi}\int d^3x e^{-3\varphi/2}e \e{^{(0)0}}
\end{equation}

Following the same procedure used in Sec.~\ref{11122018d}, we find that 
\begin{equation}
E_{\Lambda}=\frac{c^4}{6G}\Lambda a^3r^3e^{-3\varphi/2}, \label{13122018a}
\end{equation}
\begin{equation}
E_{\Lambda}=\frac{c^4}{2G}\Lambda a^3e^{-3\varphi/2}I_1, \label{13122018b}
\end{equation}
\begin{equation}
E_{\Lambda}=\frac{c^4}{2G}\Lambda a^3e^{-3\varphi/2}I_3, \label{13122018c}
\end{equation}
for $k=0,1,-1$, respectively; the integrals $I_1$ and $I_3$ are given by Eqs.~(\ref{12122018f}) and (\ref{12122018i}).

All the energies presented here are related to each other through
\begin{equation}
E=E_g+E_M+E_{\Lambda}, \label{13122018d}
\end{equation}
where
\begin{equation}
E_M=\int d^3x ee^{-5\varphi/2}\theta^{0(0)} \label{13122018e}
\end{equation}
cannot be calculated directly unless we solve the field equations.

At this point, it is important to emphasize that all of the equations in Sec.~\ref{11122018c}, \ref{11122018d}, and \ref{12122018l}, except for Eq.~(\ref{13122018e}), are independent of the coupling prescription for the matter field ($\call_M$). This will not be the case for the next section.

\subsection{Nonequivalent solution} \label{20122018e}
Let us find an example of a solution of Eq.~(\ref{12042018b}) that is not equivalent to the TEGR. In doing so, we assume that in the frame ($\e{_c^\lambda},\varphi\neq 0$) the metric is given by Eq.~(\ref{27092018c}) and  $u^{\mu}=\e{_{(0)}^\mu}=(1,0,0,0)$. So, we are assuming that the spacetime looks  both homogeneous and isotropic in a frame with a nontrivial Weyl field.

Applying these assumptions and Eqs.~(\ref{27092018d})-(\ref{09122018d}) to Eq.~(\ref{11072017a}), and then substituting the result into Eq.~(\ref{12042018b}), we finally arrive (after some manipulation) at two independent equations:
\begin{equation}
3\frac{\dot{a}^2}{a^2}+3\frac{k}{a^2}-\Lambda e^{-\varphi}-3\dot{\varphi}\frac{\dot{a}}{a}+\frac{3}{4}\dot{\varphi}^2+\chi\rho e^{-\varphi}=0, \label{25092018a}
\end{equation}
\begin{equation}
2\frac{\ddot{a}}{a}+\frac{\dot{a}^2}{a^2}+\frac{k}{a^2}-\Lambda e^{-\varphi}-2\dot{\varphi}\frac{\dot{a}}{a}+\frac{1}{4}\dot{\varphi}^2-\ddot{\varphi}-\chi p e^{-\varphi}=0. \label{25092018b}
\end{equation}
Alternatively, one may substitute Eqs.~(\ref{27092018d})-(\ref{09122018d}) directly into Eq.~(\ref{16112018d}) [Recall that $\superpotential{^a^b^c}$ is given by Eqs.~(\ref{20072017a})-(\ref{02112018a})] to obtain these equations. Notice that, since the universe is homogeneous and isotropic, we take $\rho=\rho(t)$ and $p=p(t)$ (keep in mind that they do not change under WT).

To find the conservation of energy of the matter field, let us multiply Eq.~(\ref{25092018a}) by $e^{\varphi}/\chi$, derive it and add the result to Eq.~(\ref{25092018b}) multiplied by $-3\dot{a}e^{\varphi}/(\chi a)$. The result of this calculation is the equation
\begin{align}
\dot{\rho}-\frac{9e^{\varphi}}{\chi}\frac{\dot{a}^3}{a^3}+\frac{12\dot{\varphi}e^{\varphi}}{\chi}\frac{\dot{a}^2}{a^2}-\frac{3\dot{\varphi}e^{\varphi}}{\chi}\frac{\ddot{a}}{a}-\frac{15\dot{\varphi}^2e^{\varphi}}{4\chi}\frac{\dot{a}}{a}+\frac{3\dot{\varphi}^3e^{\varphi}}{4\chi}
\nonumber\\
+\frac{3\dot{\varphi}e^{\varphi}}{2\chi}\ddot{\varphi}+\frac{3\Lambda}{\chi}\frac{\dot{a}}{a}+3p\frac{\dot{a}}{a}-\frac{9ke^{\varphi}}{\chi}\frac{\dot{a}}{a^3}+\frac{3k\dot{\varphi}e^{\varphi}}{\chi}\frac{1}{a^2}=0. \label{25092018c}
\end{align}
Using Eq.~(\ref{25092018b}) to eliminate $\ddot{\varphi}$ in Eq.~(\ref{25092018c}), we get
\begin{align}
\dot{\rho}-\frac{9e^{\varphi}}{\chi}\frac{\dot{a}}{a}\frac{\dot{a}^2}{a^2}+\frac{27\dot{\varphi}e^{\varphi}}{2\chi}\frac{\dot{a}^2}{a^2}-\frac{27\dot{\varphi}^2e^{\varphi}}{4\chi}\frac{\dot{a}}{a}+\frac{9\dot{\varphi}^3e^{\varphi}}{8\chi}+\frac{3\Lambda}{\chi}\frac{\dot{a}}{a}
\nonumber\\
-\frac{3\Lambda\dot{\varphi}}{2\chi}+3p\frac{\dot{a}}{a}-\frac{3p\dot{\varphi}}{2}-\frac{9k e^{\varphi}}{\chi}\frac{\dot{a}}{a^3}+\frac{9k\dot{\varphi}e^{\varphi}}{2\chi}\frac{1}{a^2}=0. \label{25092018d}
\end{align}
Isolating   $\dot{a}^2/a^2$ in Eq.~(\ref{25092018a}) and applying the result twice in the above equation to eliminate the terms with $\dot{a}^2/a^2$, we obtain (after some manipulation)
\begin{equation}
\frac{d}{dt}\left(\rho a^3 e^{-3\varphi/2}\right)+p\frac{d}{dt}\left(a^3e^{-3\varphi/2} \right)=0. \label{25092018f}
\end{equation}
This is exactly what we would have obtained if we had exchanged the scalar factor of the GR equation of energy conservation for $ae^{-\varphi/2}$, i.e., $\tilde{a}=ae^{-\varphi/2}$, where $\tilde{a}$ is the scalar factor in the teleparallel frame; this is in agreement with $\tilde{g}_{\mu\nu}=e^{-\varphi}g_{\mu\nu}$. However, the procedure adopted here will not give an equivalent solution for a nontrivial $\varphi$ because we have assumed the four-velocity $u^{\mu}=(1,0,0,0)$, rather than $u^{\mu}=e^{-\varphi/2}\tilde{u}^{\mu}=(e^{-\varphi/2},0,0,0,)$.

Equation (\ref{25092018f}) can be recast in terms of $E_M$ in the following way. Using the assumptions of this section in the matter energy-momentum tensor, given by Eq.~(\ref{11112018c}), we find that $\theta^{0(0)}=\rho e^{\varphi}$; therefore, Eq.~(\ref{13122018e}) becomes $E_M=C\rho a^3e^{-3\varphi/2}$, where $C$ is a constant given by ($C$ is  constant for a given $\rb$)
\begin{equation}
C=4\pi\Biggl\{ \begin{array}{cc}
r^3/3,&\textrm{\ for\ } k=0,
\\
I_1,&\textrm{\ for\ } k=1,
\\
I_3,&\textrm{\ for\ } k=-1. 
\end{array}\label{16122018a}
\end{equation}
 Multiplying Eq.~(\ref{25092018f}) by $C$ and using $V\equiv C a^3 e^{-3\varphi/2}$ as the three-dimensional invariant volume (under WT), we arrive at the first law of thermodynamics $dE_M/dt+pdV/dt=0$. Note that $E_M=\rho V$. Note also that, if the fluid is pressureless, then there is no exchange of energy between matter and gravity ($E_M$ is constant).

By using the equation of state $w=p/\rho$ we can integrate Eq.~(\ref{25092018f}) to obtain the solution
\begin{equation}
\rho(t)=\bar{\rho} a(t)^{-3(1+w)}e^{3(1+w)\varphi(t)/2}, \label{25092018g}
\end{equation}
where $\bar{\rho}$ is a constant. Substituting Eq.~(\ref{25092018g}) into Eq.~(\ref{25092018a}), we get
\begin{align}
3\frac{\dot{a}^2}{a^2}+3\frac{k}{a^2}-\Lambda e^{-\varphi}-3\dot{\varphi}\frac{\dot{a}}{a}+\frac{3}{4}\dot{\varphi}^2
+\chi\frac{\bar{\rho}e^{(1+3w)\varphi/2}}{a^{3(1+w)}}=0. \label{25092018h}
\end{align}
To solve this equation, let us assume that $k=\Lambda=p=0$ ($\omega=0$). In this case, we have
\begin{equation}
\rho=\frac{\bar{\rho}}{a^3} e^{3\varphi/2}, \label{26092018b}
\end{equation}
and
\begin{equation}
3\frac{\dot{a}^2}{a^2}-3\dot{\varphi}\frac{\dot{a}}{a}+\frac{3}{4}\dot{\varphi}^2+\frac{\chi\bar{\rho}}{a^3}e^{\varphi/2}=0.  \label{26092018a}
\end{equation}
Making the substitution $a(t)=b(t)e^{\varphi/2}$,  the above expression becomes
\begin{equation}
3\frac{\dot{b}^2}{b^2}+\frac{\chi\bar{\rho}}{b^3}e^{-\varphi}=0. \label{26092018c}
\end{equation}
Solving this equation and coming back to $a(t)$, we finally get the solution
\begin{equation}
a_{\pm}(t)=e^{\varphi/2}\left(C_1\pm \frac{3}{2}\sqrt{\frac{-\chi\bar{\rho}}{3}}\int e^{-\varphi/2}dt\right)^{2/3}, \label{26092018b}
\end{equation}
\begin{equation}
\rho(t)=\bar{\rho}\left(C_1\pm \frac{3}{2}\sqrt{\frac{-\chi\bar{\rho}}{3}}\int e^{-\varphi/2}dt\right)^{-2}, \label{26092018c}
\end{equation}
where $C_1$ is an integration constant. From Eqs.~(\ref{26092018b})-(\ref{26092018c}), (\ref{16122018a}), and $E_M=C\rho a^3e^{-3\varphi/2}$, we see that $\bar{\rho}=E_M(r)/(4\pi r^3/3)$. Furthermore, it is easy to check that the above equations together with Eq.~(\ref{12122018d}) implies  $E_g=-E_M$, which is in agreement with Eqs.~(\ref{13122018d}), (\ref{13122018a}), and (\ref{09122018g}).

As an application, consider the case where $\varphi=\alpha t$ with $\alpha>0$. In this case, we can recast Eqs.~(\ref{26092018b}) and (\ref{26092018c}) as
\begin{equation}
a_{\pm}(t)=\gamma\left(\beta e^{\alpha t/2}\mp 1 \right)^{2/3} e^{\alpha t/6}, \label{27092018a}
\end{equation}
\begin{equation}
\rho_{\pm}(t)=\xi\left(\beta \mp e^{-\alpha t/2} \right)^{-2}, \label{27092018b}
\end{equation}
where $\beta$ is an arbitrary constant associated with $C_1$, $\gamma\equiv\left(-3\chi\bar{\rho}/\alpha^2 \right)^{1/3}$, and  $\xi\equiv \alpha^2/(-3\chi)$. 

\begin{figure}[h]
\includegraphics[scale=0.25]{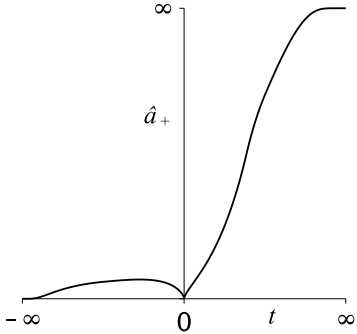}
\caption{Here, we see the evolution of the scalar factor ($\hat{a}_{+}\equiv a_{+}/\gamma$) for Eq.~(\ref{27092018a}) with the upper sign and $\beta=1$.  }
\label{13112018a}
\end{figure}

\begin{figure}[h]
\includegraphics[scale=0.25]{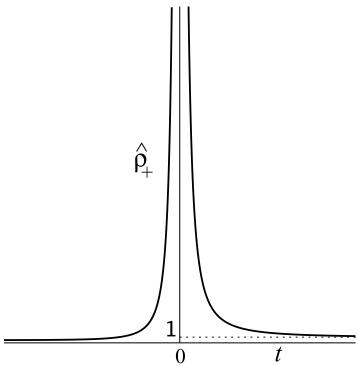}
\caption{This plot shows the behavior of $\hat{\rho}_{+}\equiv \rho_{+}/\xi$ as a function of time as given by Eq.~(\ref{27092018b}) with $\beta=1$.  Note that $\hat{\rho}_{+} \to 1$ as $t\to\infty$. }
\label{13112018b}
\end{figure}

Let us analyze first the case of the upper sign. It is clear in the equations above that, for $\beta>0$ (there is no need to assume otherwise), we must separate the case $t<0$ from $t>0$.  In the former case, we see that both the scalar factor and the energy density go to $0$ as $t\to -\infty$, see Figs.~\ref{13112018a} and \ref{13112018b}. Taking  $\beta=1$, we see that   $\rho$ diverges at $t=0$, while the universe collapses at this moment. On the other hand, for $t>0$, we have $\rho \to \infty$ while $a=0$ at the beginning. Then the universe expands indefinitely while $\rho$ goes to $\xi$.

\begin{figure}[h]
\includegraphics[scale=0.25]{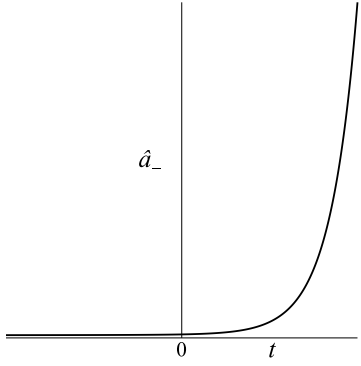}
\caption{Here, we have the behavior of $\hat{a}_{-}\equiv a_{-}/\gamma$ as a function of $t$ [given by Eq.~(\ref{27092018a})]; we have also set $\beta=1$.  }
\label{13112018c}
\end{figure}

\begin{figure}[h]
\includegraphics[scale=0.25]{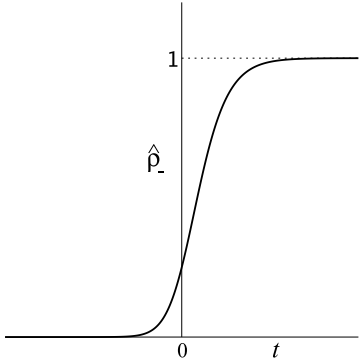}
\caption{Evolution of $\hat{\rho}_{-}\equiv \rho_{-}/\xi$ as a function of $t$ with $\beta=1$.  }
\label{13112018d}
\end{figure}

For the lower sign case, both $a_{-}$ and $\rho_{-}$ go to zero as $t\to -\infty$ and grow with $t$. The main qualitative difference between them is that $\rho_{-} \to \xi$ (taking $\beta=1$) as $t\to\infty$, while $a_{-}\to\infty$ (see Figs.~\ref{13112018c} and \ref{13112018d}).

Notice that the assumptions made at the beginning of this section are not all that obvious. Why should they hold in a frame with a nontrivial Weyl field? After all, once you change the frame by means of a WT, the new metric may not be homogeneous or isotropic anymore. Only experiments or a possible new principle could justify these choices in a particular frame. Note that it is exactly the choice of the frame where these assumptions hold that breaks the equivalence with the TEGR. In other words, if we chose these assumptions to hold in the teleparallel frame ($\tensor{\tilde{e}}{_a^\mu},\tilde{\varphi}=0$), we will get a set of solutions that are equivalent to the solution of the TEGR; However, if we make these assumptions in a frame with a nontrivial $\varphi$, there will be no guarantee that the set of solutions obtained will be equivalent to TEGR. It may turn out that the teleparallel frame is the frame that one should make these assumptions. Nonetheless, this arbitrariness should not be seen as a problem because it increases   the chances  of finding a conformal teleparallel model that  fits the experimental data.

Since $\rho$ is an invariant, the fact that Eq.~(\ref{26092018c}) is not the same (for $\varphi \neq 0$) as the one we would obtain in GR (TEGR) for the case $k=\Lambda=p=0$, that is $\rho\sim 1/t^2$ (the Einstein-de Sitter model), shows that this solution is not equivalent to that of GR (TEGR).

\section{Discussion}\label{14122017d}
In this paper we have seen how Weyl geometry can be used to manage teleparallel theories with a scalar field, mainly those theories that possess conformal invariance (translated here as invariance under WT). In particular, we have constructed a scalar $\calt$ that transforms like $\tilde{\calt}=e^{-2\theta}\calt$ under WT, regardless of the parameters of the theory.  Since all the models constructed out of $\calt$ are equivalent to a certain teleparallel model with a scalar field, it has become clear that it is easier to deal with conformal teleparallel gravity in an integrable Weyl geometry.
 
As an example, we have dealt with a model that becomes the TEGR when the Weyl field vanishes. It has been proved that this model corresponds to the model in Ref.~\cite{doi:10.1142/S0217732317501139}, except for the coupling prescription: the coupling prescription used here allows any kind of matter fields that do not depend on the affine connection to be coupled to the action, while the one in Ref.~\cite{doi:10.1142/S0217732317501139} is limited to the cases where the EMT of matter fields are traceless. In terms of the perfect-fluid case, this limitation translates to  $w=1/3$. Nonetheless, both models have the same vacuum solutions and also possess an arbitrariness with respect to the choice of the frame ($e{_a}$, $\varphi$)  where the boundary conditions must hold. As a consequence of this arbitrariness, we have two different types of solutions, namely, the ones whose ordinary boundary conditions are applied to the teleparallel frame, ensuring their equivalence with the TEGR ones (called equivalent solutions), and those that are not (nonequivalent solution). The nonequivalent solutions may become equivalent to the TEGR ones for a particular case of the scalar field, such as Eq. (\ref{22082017h}). But, in general, this will not be the case. All the solutions that we have obtained here, except for the one in Sec.~\ref{23122018a}, are nonequivalent solutions.

With respect to the $pp$-wave solution we have found that, if $\varphi$ has the specific form given by Eq.~(\ref{22082017h}), the $pp$-waves are the same as those of GR. It was also possible to find the solution for the general case $\varphi=\varphi(u,v,x,y)$ [Eqs. (\ref{8092017c}), (\ref{8092017e}) and (\ref{02112017a})]. Another example of a vacuum solution is the spherically symmetric one. In this case, we have obtained two type of solutions. The first one, given in Sec.~\ref{23122018c}, is not the Schwarzschild solution (unless $\varphi$ vanishes). The second one, on the other hand, is equivalent to the Schwarzschild solution and is written in a general frame where the spherical symmetry is still present, however the metric is not necessarily asymptotically flat. To exemplify the invariance of Eq.~(\ref{16112018g}) under WT, the energy-momentum vector $P^a$ has been calculated and, as expected, the result agrees with that of the TEGR.

The matter coupling prescription used here preserves the conformal symmetry, regardless of the matter field. The only restriction that has been imposed here is  that the matter Lagrangian does not depend on the affine connection $\nablab$. As a result, we are able to deal with all perfect fluids,  not only with hot matter, which is the case in Ref.~\cite{doi:10.1142/S0217732317501139}. The total and the gravitational energies of the universe have been calculated independently of the coupling with the matter field. In particular, we have found a cosmological solution that exhibits the power that conformal  teleparallel gravity might have to solve cosmological puzzles such as dark energy.

When dealing with more complex Lagrangians, we can use $f(\calt)$ rather than $f(T)$ to construct models that become equivalent to $f(T)$ whenever the Weyl field vanishes. As an example, we have $f(\calt)=\calt^2$, which is clearly invariant under WTs and, for $A=1/4$, $B=1/2$, $C=-1$,  is equivalent to the counterpart $f(T)=T^2$. Nevertheless, we do not have to limit ourselves to invariant models; a model like $f(\calt)=\calt^3$ is not invariant under WT and probably not equivalent to $f(T)=T^3$. But, it is certainly equivalent to some teleparallel model with a scalar field $\varphi$.

%\bibliography{biblio}

%

\end{document}